\documentclass[12pt,preprint]{aastex}
\input psfig.sty
\usepackage{epsf}

\def\beq{\begin{equation}}
\def\eeq{\end{equation}}
\def\bey{\begin{eqnarray}}
\def\eey{\end{eqnarray}}
\def\pppm{\rm P^3M}
\def\mpc{\,h^{-1}{\rm {Mpc}}}
\def\kpc{\,h^{-1}{\rm {kpc}}}
\def\kms{\,{\rm {km\, s^{-1}}}}

\def\br#1{{\mathbf r}_{#1}}
\def\bs#1{{\mathbf s}_{#1}}
\def\zetarr{\zeta(r_{12},r_{23},r_{31})}
\def\zetass{\zeta(s_{12},s_{23},s_{31})}
\def\scycl{(s_{12},s_{23},s_{31})}
\def\rpcycl{(r_{p12},r_{p23},r_{p31})}
\def\rppicycl{(r_{p12},r_{p23},r_{p31},\pi_{12},\pi_{13})}
\def\zetazrprp{\zeta_z(r_{p12},r_{p23},r_{p31},\pi_{12},\pi_{13})}

\def\zetasu{\zeta(s,u,v)}

\def\Qru{Q(r,u,v)}
\def\Qsu{Q_{red}(s,u,v)}

\def\Qrpu{Q_{proj}(r_{p},u,v)}

\def\Pirpu{\Pi(r_p,u,v)}
\def\nbar#1{{\bar n}({\mathbf r}_{#1})}
\def\nbas#1{{\bar n}({\mathbf s}_{#1})}
\def\xiz#1{\xi_z(r_{p#1},\pi_{#1})}
\def\xir#1{\xi(r_{#1})}
\def\xis#1{\xi(s_{#1})}
\def\wrp#1{w(r_{p#1})}
\def\mag{M_b - 5 \log h}

\def\gs{\mathrel{\raise1.16pt\hbox{$>$}\kern-7.0pt
\lower3.06pt\hbox{{$\scriptstyle \sim$}}}}
\def\ls{\mathrel{\raise1.16pt\hbox{$<$}\kern-7.0pt
\lower3.06pt\hbox{{$\scriptstyle \sim$}}}}
\def\gtsima{$\; \buildrel > \over \sim \;$}
\def\ltsima{$\; \buildrel < \over \sim \;$}
\def\prosima{$\; \buildrel \propto \over \sim \;$}
\def\gsim{\lower.5ex\hbox{\gtsima}}
\def\lsim{\lower.5ex\hbox{\ltsima}}
\def\simgt{\lower.5ex\hbox{\gtsima}}
\def\simlt{\lower.5ex\hbox{\ltsima}}
\def\simpr{\lower.5ex\hbox{\prosima}}

\begin{document}
\title { The Three-point Correlation Function of Galaxies Determined
from the 2dF Galaxy Redshift Survey} 
\author{Y.P. Jing${^{1,2}}$, G. B\"orner${^{2}}$} 
\affil{${^1}$Shanghai Astronomical Observatory, the Partner Group of
MPI f\"ur Astrophysik, \\Nandan Road 80, Shanghai 200030, China}
\affil {${^2}$Max-Planck-Institut f\"ur Astrophysik,
Karl-Schwarzschild-Strasse 1, \\ 85748 Garching, Germany}

\begin{abstract}

In a detailed analysis of the three point correlation function (3PCF)
for the 2dF Galaxy Redshift Survey we have accurately measured the
3PCF for galaxies of different luminosity. The 3PCF amplitudes [$\Qsu$
or $\Qrpu$] of the galaxies generally decrease with increasing
triangle size and increase with the shape parameter $v$, in
qualitative agreement with the predictions for the clustering of dark
matter in popular hierarchical CDM models. The 2dFGRS results agree
well with the results of Jing \& B\"orner for the Las Camapanas
Redshift Survey (LCRS), though the measurement accuracy is greatly
improved in the present study because the 2dFGRS survey is much larger
in size than the LCRS survey.  The dependence of the 3PCF on
luminosity is not significant, but there seems to be a trend for the
brightest galaxy sample to have a lower amplitude than the fainter
ones.

Comparing the measured 3PCF amplitudes [$\Qsu$ or $\Qrpu$] to the
prediction of a WMAP concordance model, we find that the measured
values are consistently lower than the predicted ones for dark
matter. This is most pronounced for the brightest galaxies (Sample I),
for which about one-half of the predicted $Q$ value provides a good
description of $\Qrpu$ for the 2dFGRS data.  For the less luminous
sample (Sample II), the $Q$ values are also smaller than in the dark
matter model on small scales, but on scales larger than $s=8 \mpc$ and
$r_p=3.25 \mpc$ they reach the model values.  Therefore, the galaxies
of sample II are unbiased tracers on linear scales, but the bright
galaxies (sample I) have a linear bias factor of $\sim 1.5$.  As for
the LCRS data, we may state that the best fit DM model gives higher
values for the 3PCF than observed. This indicates that the simple DM
models must be refined, either by using more sophisticated bias
models, or a more sophisticated combination of model parameters.
\end{abstract}


\section {Introduction}
To infer the spatial distribution of cosmic matter from the
observed distribution of galaxies is a nontrivial task. 
Big redshift catalogs of galaxies,
and numerical simulations of the dark matter clustering depending on
the cosmological model and on initial conditions, are the
observational and theoretical basis for a treatment of this
problem. The statistical properties, both of the theoretical models
and the observational catalogs, can be obtained by some powerful tool
like the n-point correlation functions \citep[][hereafter
P80]{peebles80}. The present state of the Universe is thought to have evolved
from initial conditions for the density field which are one specific
realization of a random process with the density contrast as the
random variable. A Gaussian distribution for the initial conditions,
such as is predicted by the inflationary scenario, is fully determined
by the two-point correlation function (2PCF), or its Fourier
transform, the power spectrum $P(k)$.

This connection has motivated an extensive use of the 2PCF to analyse
galaxy catalogs \citep[e.g., ][] {dp83,jmb98, hamilton02, zehavi02,
norberg02a,hawkins02}, the cosmic microwave background anisotropy
\citep[e.g.,][and the references therein]{spergel03}, and the cosmic
shear field \citep[e.g.,][and the references therein]{penetal03,bs01}.
Several constraints on theoretical models have already been derived
despite the fact that there are many ingredients to a specific model
which can be optimally adapted to the properties of a given galaxy
sample. The cosmological parameters, the initial power spectrum of the
DM component and the bias, i.e.  the difference in the clustering of
galaxies and DM particles, can all be adjusted to some extent.

The three-point correlation function (3PCF) $\zetarr$ characterizes
the clustering of galaxies in further detail (P80), and can provide
additional constraints for cosmogonic models.  The 3PCF is zero for a
Gaussian field, but during the time evolution of the density
perturbations the distribution develops non-Gaussian properties. These
can be measured by the 3PCF, or equivalently its Fourier-transformed
counterpart, the bispectrum, and thus additional information on the
nature of gravity and dark matter is gained, including an additional
test of the structure formation models.

The theories based on CDM models predict that the 3PCF of galaxies
depends on the shape of the linear power spectrum
\citep{f84,jb97,scocci98,bk99} and the galaxy biasing relative to the
underlying mass \citep{detal85,gf94,mjw97,mvh97,cat98}.  The
second-order perturbation theory (PT2) predicts that the 3PCF of the
dark matter depends on the shape of the triangle formed by the three
galaxies, and on the slope of the linear power spectrum
\citep[][for an excellent review]{f84,jb97,barriga02,bcgs02}.

The determination of the 3PCF was pioneered by Peebles and his
coworkers in the 1970s. They proposed a so-called ``hierarchical" form
\beq\label{hier} 
\zetarr=Q\Bigl[\xir{12}\xir{23} +\xir{23}\xir{31}
+\xir{31}\xir{12}\Bigr] 
\eeq 
with the constant $Q\approx 1.29\pm 0.2$. This form is valid for
scales $r\ls 3\mpc$ (P80).  Subsequently the analysis of several
galaxy catalogs has supported this result. The ESO-Uppsala catalog of
galaxies \citep{lau82} was analysed by \citet{jmb91}.  The 3PCF was
also examined for the CfA, AAT and KOSS redshift samples of galaxies
\citep{p81,betal83,ej84,hfms89}. These earlier redshift samples are
too small, with $\le 2000$ galaxies, to allow a test of the
hierarchical form in redshift space. Only fits to the hierarchical
form were possible.  The $Q$ value obtained in this way from redshift
samples is around $0.6$ \citep{ej84}, much smaller than the
value extracted by Peebles and his coworkers from the Lick and Zwicky
catalogs.  Redshift distortion effects are probably responsible for
this reduction \citep{m94}.

If the density field of the galaxies $\delta_g (\mathbf x)$ is 
connected to the matter overdensity $\delta_m (\mathbf x)$ as

\beq\label{deltabias}
\delta_g = b_1 \delta_m + b_2 {\delta_m}^2\,,
\eeq
then in PT2 $P_g(k) = {b_1}^2 P_m(k) $ and
\beq\label{qbias}
Q_g = Q_m/b_1 + b_2/{b_1}^2
\eeq
for the $Q$ value of the galaxy 3PCF. Since $Q_m$ depends on the shape
of the power spectrum in PT2 that can be measured from the galaxy
power spectrum on large scales (assuming a linear bias), one may
measure the bias parameters $b_1$ and $b_2$ from the 3PCF of galaxies
on large scales.

The hierarchical form (Eq.~\ref{hier}) is purely empirical without a
solid theoretical argument supporting it. In contrast, the PT2 theory
predicts that $Q_m$ of dark matter depends on the shape of triangles
on the linear clustering scale. Even in the strongly non-linear regime
where the hierarchical form was expected to hold, the CDM models do
not seem to obey it, as demonstrated by \citet[][hereafter
JB98]{jb98}. The large sample size of the Las Campanas Redshift Survey
\citep[LCRS; ][]{shectman96} made it possible for the first time to
study the detailed dependence of the amplitude $Q_g$ of galaxies on
the shape and size of triangles. JB98 computed the 3PCFs for the LCRS
both in redshift space and in projected space. As demonstrated by
JB98, the projected 3PCF they proposed has simple relations to the
real space 3PCF.  Their results have revealed that both in redshift
space and in real space there are small, but significant deviations
from the hierarchical form.

The general dependence of the galaxy 3PCF on triangle shape and size
appeared to be in qualitative agreement with the CDM cosmogonic
models. JB98 found that a CDM model with $ \Omega_m h = 0.2 $, and an
appropriately chosen bias scheme (the Cluster-Weighted model
originally proposed in \citep[][hereafter JMB98]{jmb98}, now generally
called Halo- Occupation- Number model in the literature) meets the
constraints imposed by the LCRS data on the 2PCF and the pairwise
velocity dispersion (PVD) of the galaxies.  The real-space $Q_g$
obtained from the LCRS is, however, well described by half the mean
$Q_m$ value predicted by this best-fit CDM model. The unavoidable
conclusion is that it is difficult to find a simple model which meets
all the constraints.

In recent years, several authors have measured the 3PCF and the
bispectrum, with emphasis on the quasilinear and linear
clustering scales.  For example, for the APM galaxies
\citep{gf94,fg99}, the IRAS galaxies \citep{scoccimarro01}, and the
2dFGRS galaxies \citep{verde02}, the measurements were used to
constrain the linear, and nonlinear bias parameters $b_1$ and $b_2$
(Eq.\ref{deltabias}), by comparison with a model for the 3PCF
obtained in PT2.

For the APM galaxies the PT2 model for the 3PCF agrees well with the
APM catalog measurements on large scales \citep{fg99}, which implies $b_1\approx 1$ and $b_2\approx 0$. The bispectrum
of PSCz IRAS galaxies leads to values of
\bey\label{biras} b_1^{-1}&=&1.32 (+0.36,-0.39)\\ 
b_2^{-1}&=&1.15\pm 0.39 
\eey 
\citep{scoccimarro01} for the wavenumber $k$ in the interval $0.05
\le k \le 0.2 \,h \rm {Mpc}^{-1} $. The measurement of the bispectrum
for the 2dFGRS catalog resulted in bias parameters
\bey\label{b2df}
b_1&=&1.04 \pm 0.11\\
 b_2&=&-0.054 \pm 0.08
\eey
on scales between $5$ and $30 \mpc$ \citep{verde02}. These results
indicate that on large scales, optical galaxies (both 2dFGRS galaxies
and APM galaxies) are unbiased relative to the underlying mass
distribution, while the IRAS galaxies are an anti-biased
tracer. Furthermore the non-linear bias of the IRAS galaxies is
significantly non-zero. Combining these results with our result on the
LCRS (JB98) implies that optical galaxies are a biased tracer on small
scale, but an unbiased tracer on larger scale.

In this paper, we measure the 3PCF both in redshift and in the
projected space for the Two Degrees Fields Galaxy Redshift Survey
\citep[][2dFGRS]{colless01}.  We are motivated to investigate further
the mismatch of the 3PCF found by JB98 between the concordance CDM
model and the LCRS survey. Because the 2dFGRS covers a much larger
volume than the LCRS, we expect to measure the 3PCF more accurately
especially on large scales. Therefore we attempt to find out, if there
exists a transition where the 3PCF gradually approaches the unbiased
prediction of the concordance CDM model on large scales
\citep{fg99,verde02} from half of the CDM prediction on small scales
(JB98). The results of \citet{fg99} and \citet{verde02} apparently
imply a high normalization $\sigma_8\approx 1$ for the primordial
fluctuation ( $\sigma_8$ is the linear rms density fluctuation at the
present in a sphere of radius $8\mpc$), while some observations,
e.g. the PVD of galaxies, the abundance of clusters of galaxies,
clearly prefer a smaller value of $\sigma_8\approx 0.7$ for the
concordance LCDM model
\citep[e.g.][]{bahcall2000,lahav2002,vandenbosch03,yang03b}. This
apparent conflict also motivates us to examine the 3PCF more carefully
on quasilinear scales which can be explored by the 2dFGRS.  Moreover,
it is well known that the clustering of galaxies depends on their
luminosity. In \citet{fg99} and \citet{verde02} galaxies are included
in a wide range of luminosity, and it is difficult to determine,
whether for some luminosity range galaxies are unbiased relative to
the mass distribution on large scales. In this paper, we will attempt
to measure the 3PCF for the first time as a function of luminosity. We
believe that these measurements of the 3PCF will provide useful
observational constraints on galaxy formation theories.

In section 2, we will describe the sample selection for the analysis,
the selection effects, and the procedure of generating random and mock
samples. The statistical methods of measuring the 3PCFs are presented
in Section 3. The results of the 2dFGRS are given in Section 4, along
with a comparison with the results of the LCRS and the predictions of
the concordance model for dark matter. Our results are summarized in
Section 5.

\section{Observational sample, random sample, and mock catalogs}
We select data for our analysis from the 100k public release
\footnote{Available at http://www.mso.anu.edu.au/2dFGRS} of the 2dFGRS
\citep[][; hereafter C01]{colless01}. The survey covers two declination
strips, one in the Southern Galactic Pole (SGP) and other in the
Northern Galactic Pole(NGP), and 99 random fields in the southern
galactic cap. In this paper, only galaxies in the two strips are
considered. Further criteria for the inclusion of galaxies in our
analysis are that they are within the redshift range of $0.02<z<0.25$,
have the redshift measurement quality $Q\ge 3$, and are in regions
with redshift sampling completeness $R({\mathbf \theta})$ better
than 0.1 (where ${\mathbf \theta}$ is a sky position). The redshift
range restriction is imposed so that the clustering statistics are
less affected by the galaxies in the local supercluster, and by the
sparse sampling at high redshift. The redshift quality restriction is
imposed so that only galaxies with reliable redshifts are used in our
analysis. An additional reason is that the redshift completeness mask
provided by the survey team, which is used in our analysis, is
constructed for the redshift catalog of $Q\ge 3$. The last restriction
is imposed in order to eliminate galaxies in the fields for which the
field redshift completeness $c_{F}$ is less than 70 percent (see C01
about the difference between $R({\mathbf \theta})$ and $c_F$). These
fields are (or will be) re-observed,and have not been included in
computing the redshift mask map $R({\mathbf \theta})$. Finally,
there are a total of 69655 galaxies satisfying our selection criteria,
30447 in the NGP strip and 39208 in the SGP strip.

It is well known that the two-point clustering of galaxies depends on
the luminosity \citep{xia87, boerner91, loveday95,norberg02a},
and the luminosity dependence is an important constraint on galaxy
formation models \citep{kauffmann97,kauffmann99,benson00,
yang03a}. We take advantage of the size of the 2dFGRS to carry out a
first study of the luminosity dependence of the three point
correlation function. The galaxies are divided into three classes;
luminous galaxies with absolute magnitude $M_b \le
M_b^{\star}=-19.66+5\log h$, faint galaxies with $M_b>-18.5+5\log h$,
and typical galaxies with luminosity in between, where $
M_b^{\star}$ is the characteristic luminosity of the Schechter
function in the $b_J$ band \citep{norberg02b}, and $h$ is the Hubble
constant in units of $100\kms{\rm Mpc}^{-1}$. We will also do the analysis
for galaxies with $M_b\le -18.5+5\log h$ in order to compare 
the results with the
previous study of the Las Campanas Redshift Survey \citep{jb98}.  The
details of the subsamples studied in this paper are given in Table 1.
For computing the absolute magnitude, we have used the k-correction
and luminosity evolution model of \citet[][${\rm k+e}$
model]{norberg02b}, i.e., the absolute magnitude is in the rest
frame $b_j$ band at $z=0$. We assume a cosmological model with the
density parameter $\Omega_0=0.3$ and the cosmological constant
$\lambda_0=0.7$ throughout this paper.

A detailed account for the observational selection effects 
has been released
with the catalog by the survey team (C01). The limiting magnitude
changes slightly across the survey region due to further magnitude
calibrations that were carried out after the target galaxies had been
selected for the redshift measurement. This observational effect is
documented in the magnitude limit mask $b_J^{\rm lim} (\mathbf \theta)
$ (C01). The redshift sampling is far from uniform within the survey
region, and this selection effect is given by the redshift
completeness mask $R (\mathbf \theta)$. The redshift measurement
success rate also depends on the brightness of galaxies, making
fainter galaxies more incomplete in the redshift measurement. The $\mu
( \mathbf \theta)$ mask provided by the survey team is aimed to
account for the brightness-dependent incompleteness.

These observational effects can be corrected in our analysis for the
three-point correlation function through properly generating random
samples. To construct the random samples, we first select a spatial
volume that is sufficiently large to contain the survey sample. Then,
we randomly distribute points within the volume, and eliminate the
points that are out of the survey boundary. Adopting $15.0<b_J\le
b_J^{\rm lim} (\mathbf \theta)$ for the magnitude limits\footnote{We
assume that the brighter magnitude limit for the survey is 15.0. This
is a reasonable value for the survey, but our results are insensitive
to the choice of this value.} of the survey in the direction ${\mathbf
\theta}$, we select random points according to the luminosity
function of the 2dFGRS and the $k+e$ model for the k-correction and
luminosity evolution \citep{norberg02a}, and assign to each point an
apparent magnitude (and an absolute magnitude). This unclustered
sample is a random sample for the 2dFGRS photometric catalog. Then we
implement the magnitude-dependent redshift selection effect 
according to C01. We keep random points of magnitude $m$ in
the direction ${\mathbf \theta}$ at a sampling rate $S({\mathbf \theta},m)$
which reads as [Eq.(11) of C01],
\beq
S({\mathbf \theta},m)=\frac{N_p({\mathbf \theta})}{N_e({\mathbf \theta})}
R({\mathbf \theta}) c_z[m,\mu({\mathbf\theta})] 
\eeq
where $N_p({\mathbf \theta})$ is the number of parent catalog galaxies
in the sector ${\mathbf \theta}$ and $N_e({\mathbf \theta})$ is the
number of galaxies which are expected  to have measured redshifts for
given $\mu({\mathbf \theta})$. The ratio
$N_e({\mathbf\theta})/N_p({\mathbf\theta})$ is actually the field
completeness $c_F(\mathbf \theta)$ defined by C01 which we compute
according to their Eq.(7) \citep[see also ][]{norberg02b}. The function
$R({\mathbf \theta})$ is given by the redshift completeness mask and
$c_z[m,\mu({\mathbf\theta})]$ can be easily computed from the $\mu$
mask [eq.(5) of C01]. We have used the corrected value $0.5\ln (10)$
for the $\alpha$ parameter in the power-law galaxy count model
according to the Web page of the 2dFGRS.

\begin{figure}
\epsscale{1.0} \plotone{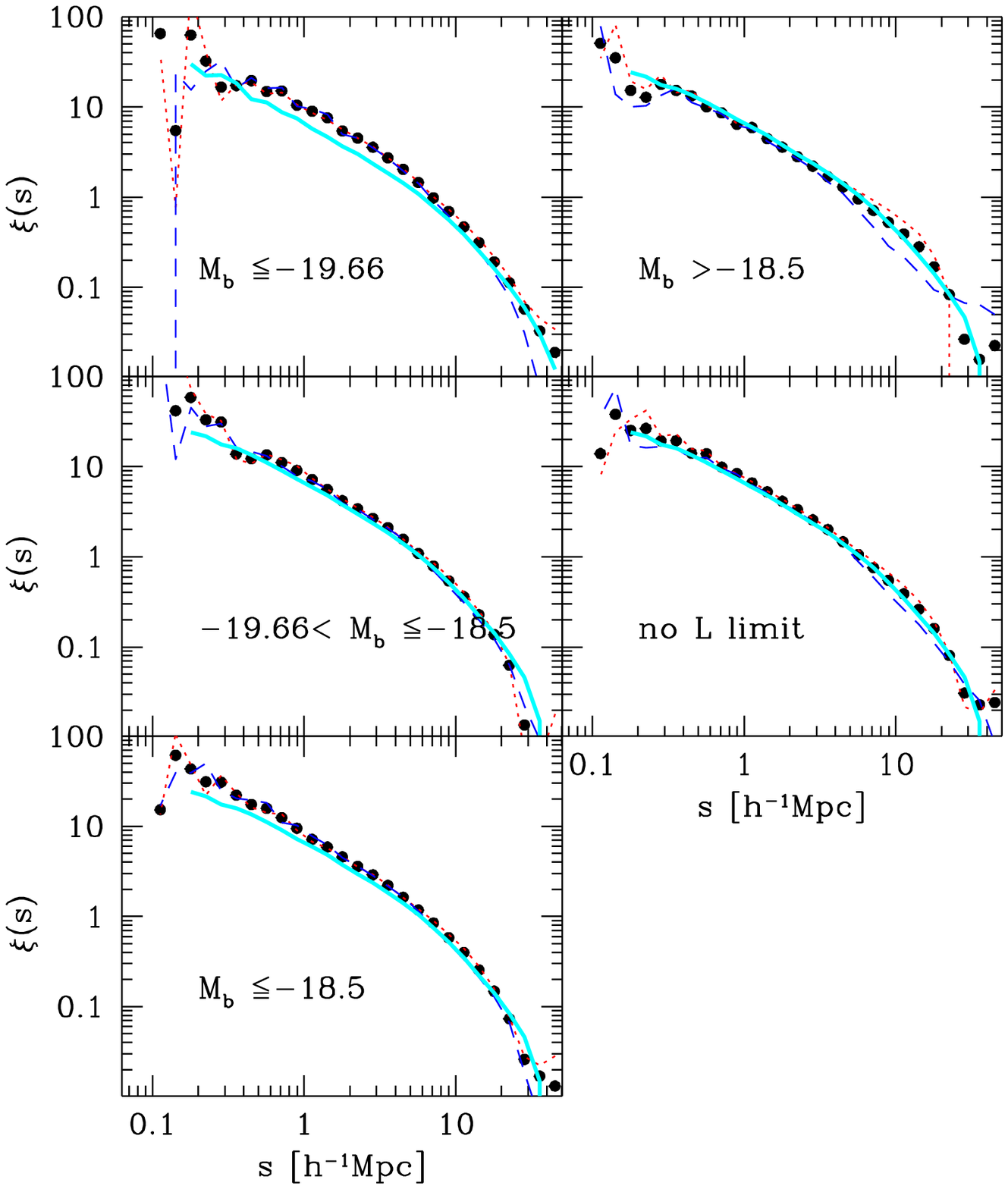}
\caption{The redshift two-point correlation function of galaxies with
different luminosity measured from the 2dFGRS catalog. The symbols
with error bars are for the whole catalog, the dotted lines are for
the southern subsample, and the dashed lines are for the northern
subsample. The errors are estimated by the bootstrap resampling
method. The luminosity ranges are indicated at each panel. The thick
solid lines are the simulation predictions for the redhsift two-point
correlation of dark matter in the WMAP concordance CDM model at
redshift $z=0.13$.}
\label{xis}\end{figure}

\begin{figure}
\epsscale{1.0} \plotone{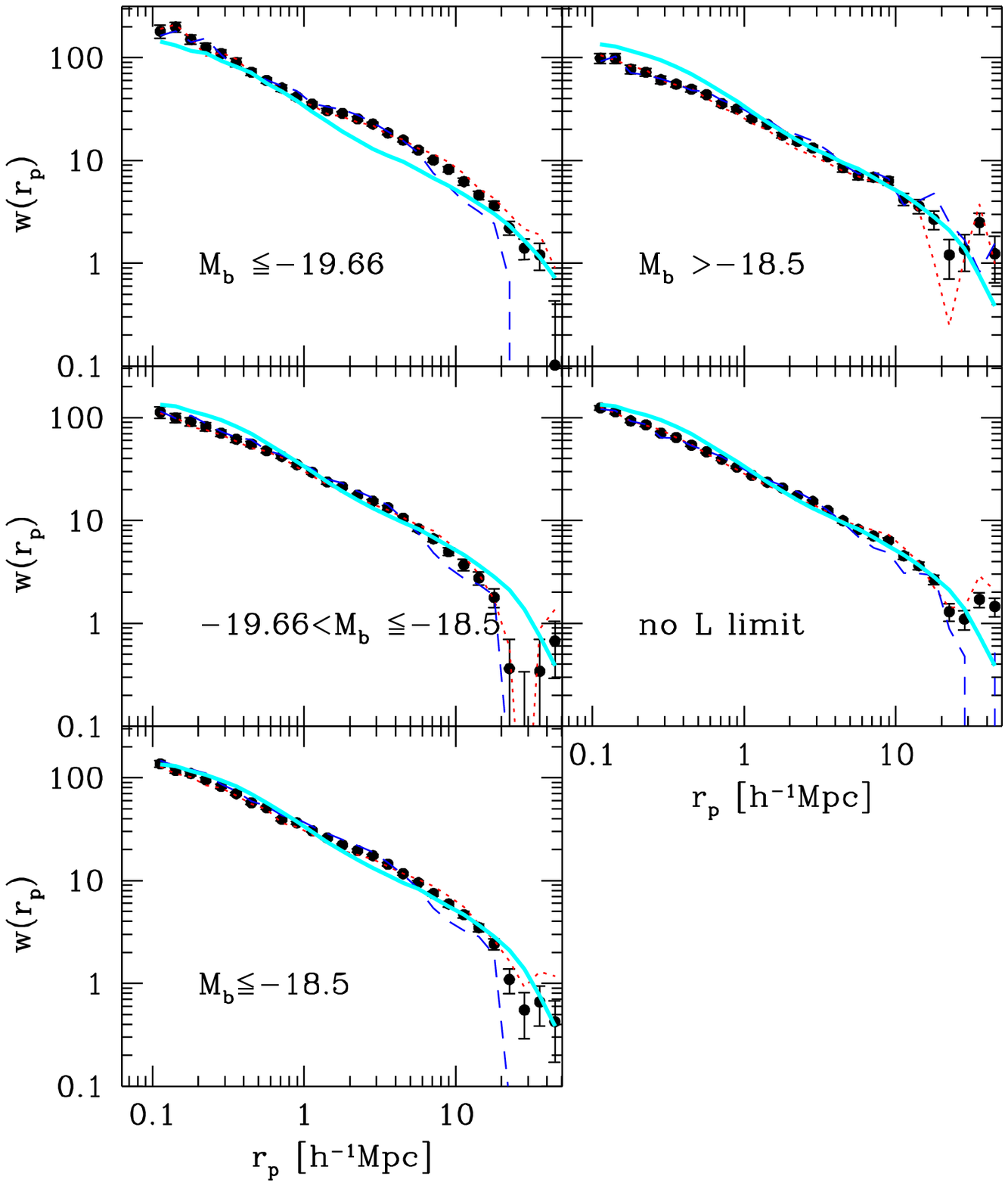}
\caption{The projected two-point correlation function of galaxies with
different luminosity measured from the 2dFGRS catalog. The symbols
with error bars are for the whole catalog, the dotted lines are for
the southern subsample, and the dashed lines are for the northern
subsample. The errors are estimated by the bootstrap resampling
method. The luminosity ranges are indicated at each panel. The thick
solid lines are the simulation predictions for the projected two-point
correlation of dark matter in the WMAP concordance CDM model at
redshift $z=0.13$. }
\label{wrpfig}\end{figure}

We have checked the random samples carefully by reproducing the
angular distribution, mean redshift distribution, and especially the
two-point statistics of clustering of the observed catalog. It is
known that the two-point correlation function measured from galaxy
catalogs on large scale is sensitive to the details of corrections for
the above selection effects. We have estimated the redshift and
projected correlation functions by the same method as in JMB98 for the
Las Campanas Redshift Survey. The two-point correlation functions are
shown in Figure \ref{xis} and Figure \ref{wrpfig}, and can be compared
with the results of the 2dFGRS team for the clustering of galaxies
\citep[e.g.][]{hawkins02,norberg02a}. In addition to the broad
agreement with their results, even the subtle difference between the
north and south caps (the clustering on large scales is slightly
larger in the southern cap than in the northern cap), and the
luminosity dependence of the clustering, is well reproduced in our
analysis.

We did not take into account in our analysis the fiber collision
effect that two galaxies closer than $\sim 30$ arcsec cannot be
assigned fibers simultaneously in one spectroscopic observation. Thus
one of them will not have a redshift observation if no re-observation
is arranged. This effect reduces the real space (or projected)
two-point correlation function at small separations. With LCRS, JMB98
estimated the effect to lead to a 15 percent reduction in the
two-point correlation function at projected separations $100 \kpc$ and
to a less than 5 percent reduction at separations larger than
$400\kpc$. This effect is smaller (10 percent reduction at separations
$100\kpc$) in the 2dFGRS \citep{hawkins02}, because the limiting fiber
separation is slightly smaller (30 arc sec in the 2dFGRS vs 55 arcsec
in the LCRS), and one field may be observed more than once in the
2dFGRS observation strategy.  JB98 have examined the fiber collision
effect on their measurement of the three-point correlation function of
the LCRS. They found that the effect reduces the real space
(projected) three-point correlation function at small separation, but
changes little the normalized three-point correlations functions $Q$
that we will measure in this paper, because the effects on the
two-point CF and three-point CF are canceled out when $Q$ is
measured. Since the effect is slightly smaller in 2dFGRS in terms of
the two-point clustering, we believe that only a negligible effect on
our measurement of the normalized three-point correlations would
result.

\section{Statistical methods}
We measure the three-point correlation functions for the galaxies in
the 2dFGRS following the method of JB98. By
definition, the joint probability $dP_{123}$ of finding one object
simultaneously in each of the three volume elements $d\br{1}$,
$d\br{2}$ and $d\br{3}$ at positions $\br{1}$, $\br{2}$ and $\br{3}$
respectively, is as follows (P80):
\beq\label{3pcfdf}
dP_{123}=\nbar{1}\nbar{2}\nbar{3}\bigl[1+\xir{12}+\xir{23}+\xir{31}+
\zetarr\bigr] d\br{1} d\br{2} d\br{3} 
\eeq 
where $r_{ij}=|\br{i}-\br{j}|$, $\nbar{i}$ is the mean density of
galaxies at $\br{i}$, and $\zetarr$ is the three-point correlation
function. This definition can be applied straightforwardly to redshift
surveys of galaxies to measure the 3PCF $\zetass$ of galaxies in
redshift space (at this point we neglect the anisotropy induced by the
redshift distortion which will be considered later). Here and below we
use $\mathbf r$ to denote the real space and $\mathbf s$ the redshift
space.

The 3PCF of galaxies can be measured from the counts of different
triplets (P80). Four types of distinct triplets with triangles in the
range ($s_{12}\pm 1/2 \Delta s_{12}$, $s_{23}\pm 1/2 \Delta s_{23}$,
and $s_{31}\pm 1/2 \Delta s_{31}$) are counted: the count $DDD\scycl$
of triplets formed by three galaxies; the count $DDR\scycl$ of
triplets formed by two galaxies and one random point; the count
$DRR\scycl$ of triplets formed by one galaxy and two random points;
the count $RRR\scycl$ of triplets formed by three random points. The
random sample of points is generated in the way described in the
previous section. Following the definition [eq(\ref{3pcfdf})], we
shall use the following estimator
\bey\label{3pcfred} 
\zetass&=&{27 RRR^2\scycl \times DDD\scycl\over
DRR^3\scycl}\nonumber\\ 
&&- {9 RRR\scycl \times DDR\scycl\over DRR^2\scycl} +2 
\eey
to measure the 3PCF of the galaxies in redshift space. The above
formula is slightly different from the estimator used by \citet{gp77}. Here we have extended the argument of \citet{h93}
for the 2PCF to the case of the 3PCF. The coefficients $27$ and $9$
are due to the fact that only {\it distinct} triplets are counted in
this paper. Since the early work of Peebles and coworkers (P80)
indicates that the 3PCF of galaxies is approximately hierarchical, it
is convenient to express the 3PCF in a normalized form
$Q_{red}\scycl$:
\beq\label{qred} 
Q_{red}\scycl ={\zetass \over \xis{12}\xis{23}
+\xis{23}\xis{31} +\xis{31}\xis{12}} \,.  
\eeq 
It is also convenient to use the variables introduced by Peebles (P80)
to describe the shape of the triangles formed by the galaxy
triplets. For a triangle with the three sides $s_{12}\le s_{23} \le
s_{31}$, $s$, $u$, and $v$ are defined as:
\beq\label{ruv}
s = s_{12}, \hskip1cm u={{s_{23}}\over{s_{\rm 12}}},\hskip1cm
v={{s_{31}-s_{23}}\over{s_{12}}}\,.
\eeq
Clearly, $u$ and $v$ characterize the shape and $s$ the size of a
triangle. We take equal logarithmic bins for $s$ and $u$ with the bin
intervals $\Delta \log s=\Delta \log u=0.2$, and equal linear bins for
$v$ with $\Delta v=0.2$. For our analysis, we take the following
ranges for $s$, $u$ and $v$: $0.63\le s\le 10\mpc$ ($6$ bins); $1\le
u\le4$ (3 bins); and $0\le v\le 1$ (5 bins).

As in JB98, we have generalized the ordinary linked-list technique of
$\pppm$ simulations \citep{he80} to spherical coordinates to count the
triplets. The linked-list cells are specified by the spherical
coordinates, i.e. the right ascension $\alpha$, the declination
$\delta$ and the distance $s$.  With this short-range searching
technique, we can avoid the triplets out of the range specified thus
making counting triplets very efficient. Because the triplet count
$RRR$ is proportional to the third power of the number density of
random points, the count within a fixed range of triangles would vary
significantly among different luminosity subsamples if the number of
random points is fixed, since the volumes covered by different
subsamples are very different.  We want to have random samples such
that the random counts and the cross counts are as big as possible in
order to suppress any uncertainty from the limited number of random
points. Therefore, since the CPU time for counting triplets is
approximately proportional to the total count of triplets in our
linked list method, we choose the number of random points as large as
possible for the computations of $RRR$, $DRR$, or $RDD$ under the
condition that each computation is finished in $\ls 24$ CPU hours on a
Pentium IV 2.2 Ghz PC. The number of random points ranges from
$40,000$ (for Sample IV) to $120,000$ (for Sample I) when computing
$RRR$, and increases to $600,000$ (for Sample I) when computing
$RDD$. The counts $RRR$ for small triangles ($s< 1\mpc$) could still
be small, and therefore we have recalculated the counts $RRR$ for
$s_{31}\le 4\mpc$ by generating a random sample 10 times larger, so as
to ensure that the counts $RRR$ are at least $\sim 300$ for the
triangle configurations of interest.  We scaled these counts properly
when we determined the three-point correlation function through
equation (\ref{3pcfred}).  The uncertainty caused by the number of
random points is negligible compared to the sampling errors of the
observational sample.

The 3PCF in  redshift space $\Qsu$ depends both on the real space
distribution of galaxies and on their peculiar motions. Although this
information contained in $\Qsu$ is also useful for the study of the
large scale structures (see \S 4), it is apparent that $\Qsu$ is
different from  $\Qru$ in real space. In analogy with the analysis
for the two-point correlation function, we have determined the
projected three-point correlation function $\Pi\rpcycl$. We
define the redshift space three-point correlation function 
$\zetazrprp$ through:
\bey\label{3pcfzdf}
dP^z_{123}&=&\nbas{1}\nbas{2}\nbas{3}
\bigl[1+\xiz{12}+\xiz{23}+\xiz{31}\nonumber\\
&&+
\zetazrprp\bigr] d\bs{1} d\bs{2} d\bs{3} 
\eey 
where $dP^z_{123}$ is the joint probability of finding one object
simultaneously in each of the three volume elements $d\bs{1}$,
$d\bs{2}$ and $d\bs{3}$ at positions $\bs{1}$, $\bs{2}$ and $\bs{3}$;
$\xiz{}$ is the redshift space two-point correlation function;
$r_{pij}$ and $\pi_{ij}$ are the separations of objects $i$ and $j$
perpendicular to and along the line-of-sight respectively.
The projected 3PCF $\Pi\rpcycl$ is then defined as:
\beq\label{3pcfproj1}
\Pi\rpcycl = \int \zetazrprp d\pi_{12} d\pi_{23}
\eeq
Because the total amount of triplets along the line-of-sight is not
distorted by the peculiar motions, the projected 3PCF $\Pi\rpcycl$ is
related to the 3PCF in real space $\zetarr$ :
\beq\label{3pcfproj2}
\Pi\rpcycl =\int \zeta(\sqrt{r_{p12}^2+y_{12}^2},\sqrt{r_{p23}^2+y_{23}^2},
\sqrt{r_{p31}^2+(y_{12}+y_{23})^2})dy_{12} dy_{23}\,.
\eeq

Similarly as for $\zetass$, We measure $\zetazrprp$ 
similarly to $\zetass$ by
counting the numbers of triplets $DDD\rppicycl$, $DRR\rppicycl$,
$RDD\rppicycl$ and $RRR\rppicycl$ formed by galaxies and/or random points
with the projected separations $r_{p12}$, $r_{p23}$, and $r_{p31}$ and
radial separations $\pi_{12}$ and $\pi_{23}$.  We will use $r_p$, $u$
and $v$:
\beq\label{rpuv}
r_p = r_{p12}, \hskip1cm u={{r_{p23}}\over{r_{p12}}},\hskip1cm
v={{r_{p31}-r_{p23}}\over{r_{p12}}}\,.
\eeq
to quantify a triangle with $r_{p12}\le r_{p23}\le r_{p31}$ on the
plane perpendicular to the line of sight.
Equal logarithmic bins of intervals $\Delta \log
r_p=\Delta \log u=0.2$ are taken for $r_p$ and $u$, and equal linear
bins of $\Delta v=0.2$ for $v$. The same ranges of $u$ and $v$ are
used as for $\zetasu$, but $r_p$ is from $0.128\mpc$ to $4\mpc$ (7
bins). The radial separations $\pi_{12}$ and $\pi_{23}$ are from
$-25\mpc$ to $25\mpc$ with a bin size of $1\mpc$.  The projected
3PCF is estimated by summing up $\zeta_z(r_p, u,
v,\pi^i_{12},\pi^j_{23})$ at different radial bins
($\pi^i_{12},\pi^j_{23}$):
\beq\label{Pistat}
\Pirpu= \sum_{i,j} \zeta_z(r_p, u, v,\pi^i_{12},\pi^j_{23})
 \Delta\pi^i_{12} \Delta\pi^j_{23}
\eeq
and normalized as
\beq\label{qproj} 
\Qrpu ={\Pi(r_p,u,v) \over \wrp{12}\wrp{23}
+\wrp{23}\wrp{31} +\wrp{31}\wrp{12}} \,.  
\eeq 
where $\wrp{}$ is the projected two-point correlation function
\citep[][JMB98]{dp83}
\beq\label{wrp}
\wrp{} = \sum_{i}\xi_z(r_p,\pi^i)\Delta \pi^i
\eeq
An interesting property of the projected 3PCF is that if the
three-point correlation function is of the hierarchical form, 
the normalized function $\Qrpu$ is not only a constant but also equal
to $Q$. Therefore the measurement of $\Qrpu$ can be used to test the
hierarchical form which was proposed mainly based on the analysis
of angular catalogs.

\citet{jb98} have used N-body simulations to test the
statistical methods for the LCRS, and found that the results obtained
are unbiased. Since the 2dFGRS is constructed in a similar way to the
LCRS and the survey area is larger, the above methods should also
yield unbiased results for the 2dFGRS.

The error bars of $Q$ are estimated by the bootstrap method
\citep{barrow84, mjb92}. We have also used the mock samples of dark
matter particles in \S 4 to estimate the error bars. We find that the
error bars from these two methods agree within a factor of 2. Here we 
adopt the bootstrap error for the measurement of $Q$, since we do not
input a luminosity-dependent bias for mock samples.

As in the analysis of the 2PCF, the estimates of the 3PCF given by
Eq.(\ref{3pcfred}) are correlated on different scales. This point
should be taken into account when the measured 3PCF is compared with
model predictions. Recently, there are new techniques developed to
tackle this important issue in the context of the 2PCF or the power spectrum,
e.g., \citet{thy02} and \citet{ms02} using the Karhunen-Lo\`eve
eigenmode analysis, and \citet{ff00} and \citet{zjf01} using the
wavelet analysis. It remains an important task to study if these
methods can be extended to obtain a decorrelated 3PCF.

\section{Results of the 2dFGRS Analysis}

\subsection{The 3PCF of the 2dFGRS catalog, and the luminosity dependence}

\begin{figure}
\epsscale{1.0} \plotone{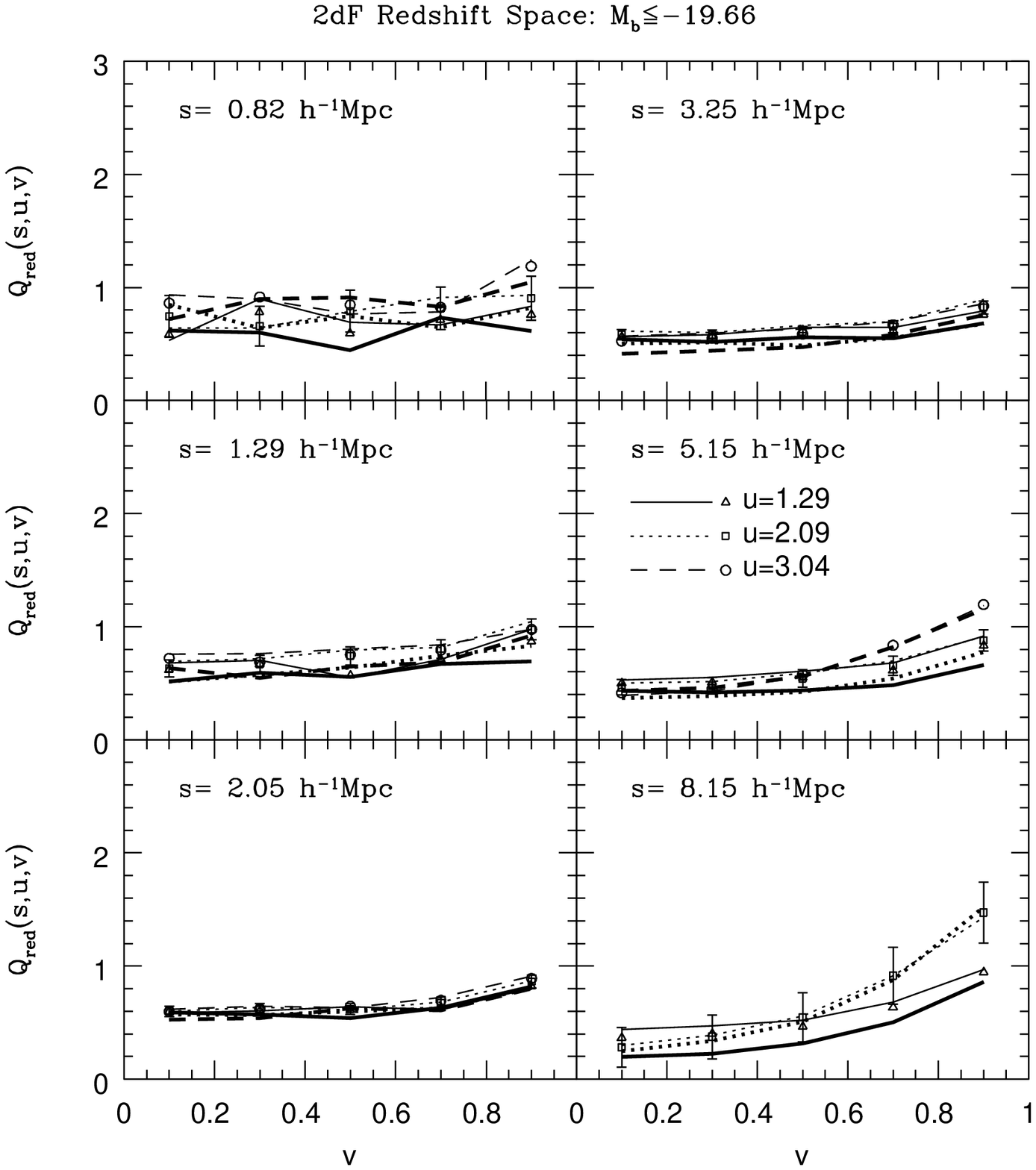}
\caption{ The normalized 3PCF in redshift space $\Qsu$ of galaxies
with luminosity $M_b-5\log h\le -19.66$ in the 2dFGRS survey. The
results for the south strip, the north strip, and the whole sample are
plotted with thick lines, thin lines, and symbols
respectively. Different lines and symbols are used for triangle
configurations of different $u$ as indicated on the figure. The errors
are estimated by the bootstrap resampling method. For clarity, the
error bars are plotted for the whole sample and $u=2$ only, but those
for the other two values of $u$ are very similar, and for north or
south strips are about $1.4$ times larger.}
\label{qred_1}\end{figure}

\begin{figure}
\epsscale{1.0} \plotone{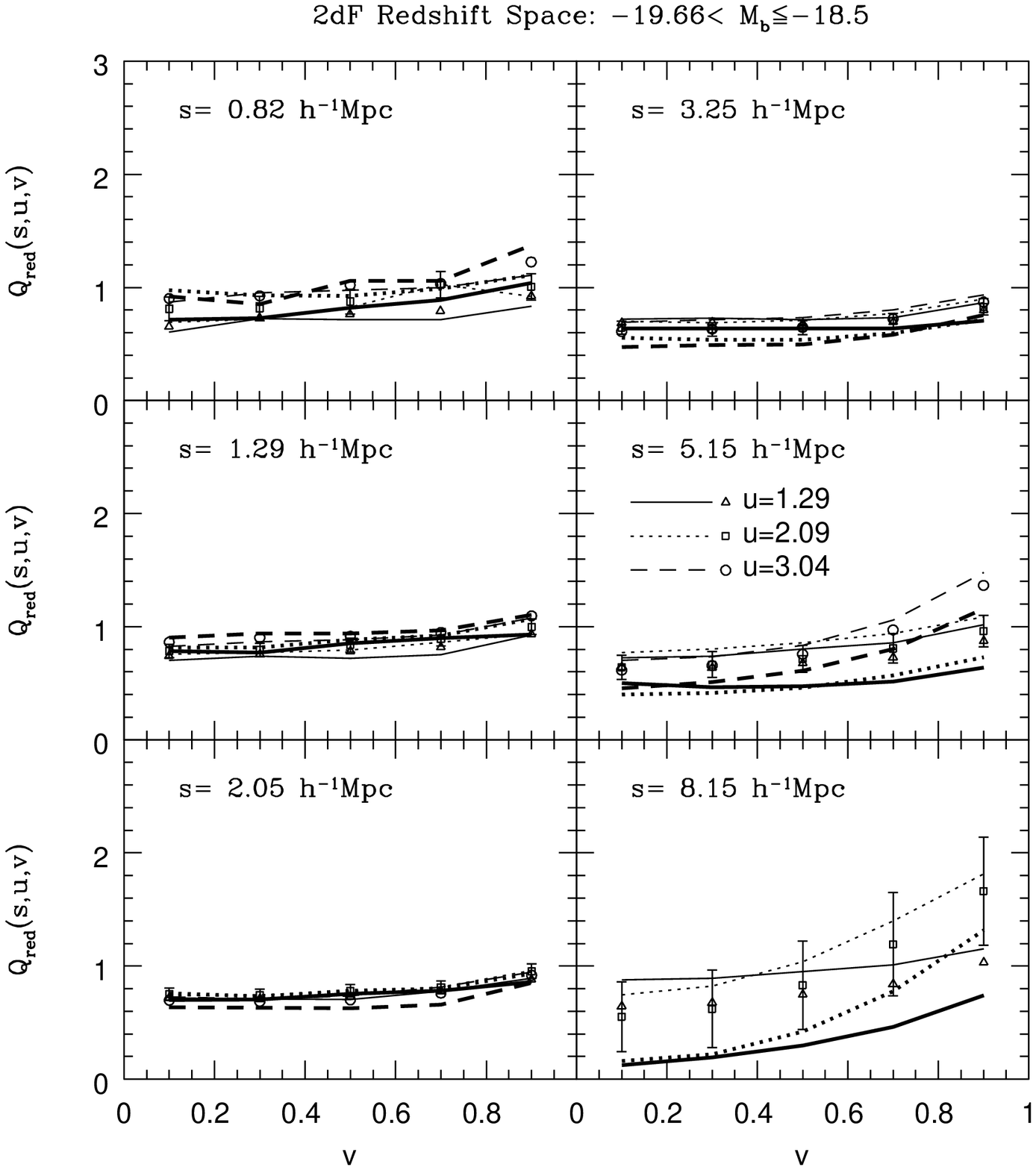}
\caption{The normalized 3PCF in redshift space $\Qsu$ of galaxies
with luminosity $-19.66<M_b-5\log h\le -18.5$ in the 2dFGRS survey.
The notations are the same as Fig.\ref{qred_1}.}
\label{qred_2}\end{figure}

\begin{figure}
\epsscale{1.0} \plotone{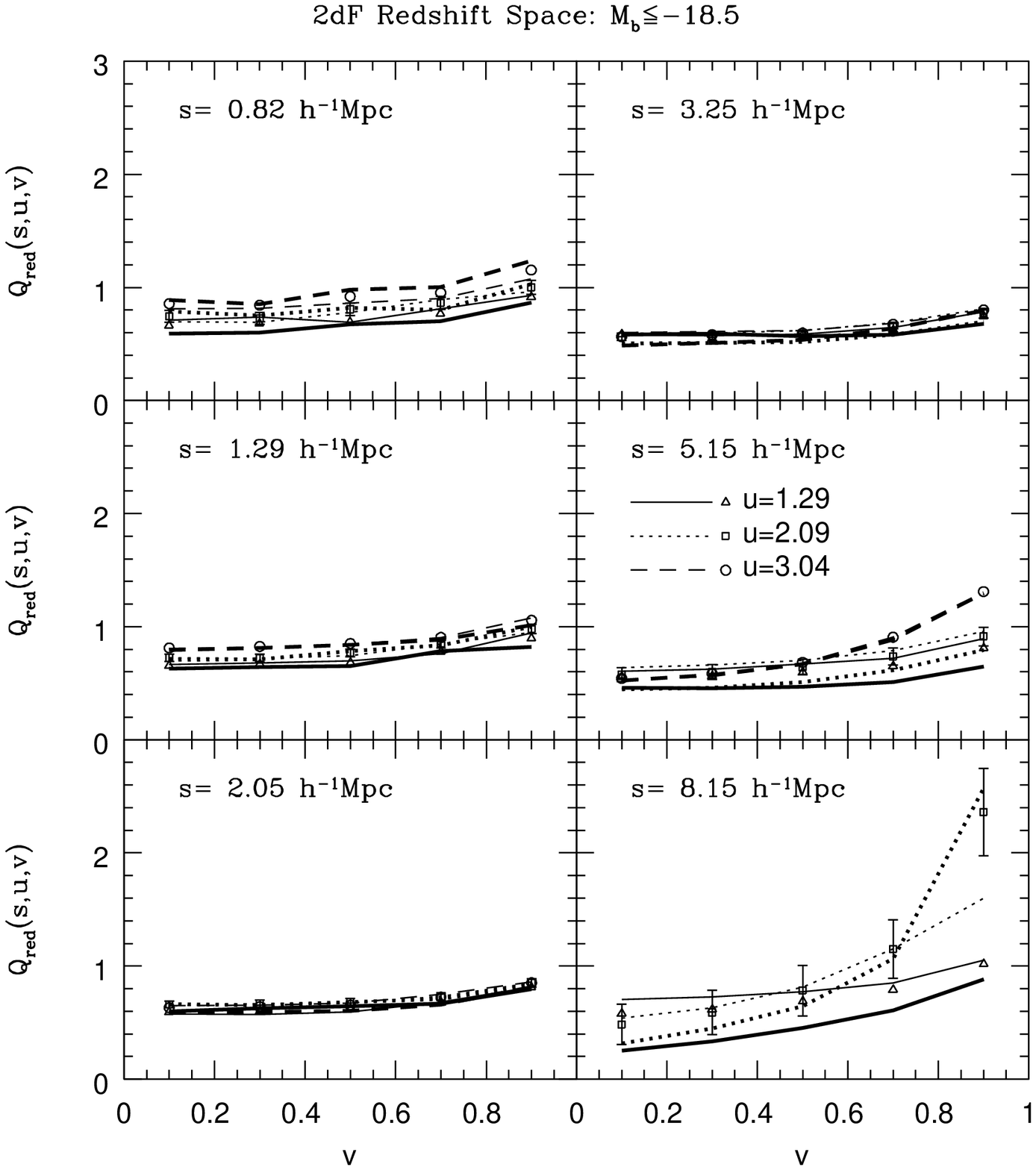}
\caption{The normalized 3PCF in redshift space $\Qsu$ of galaxies
with luminosity $M_b-5\log h\le -18.5$ in the 2dFGRS survey.
The notations are the same as Fig.\ref{qred_1}.}
\label{qred_3}\end{figure}

\begin{figure}
\epsscale{1.0} \plotone{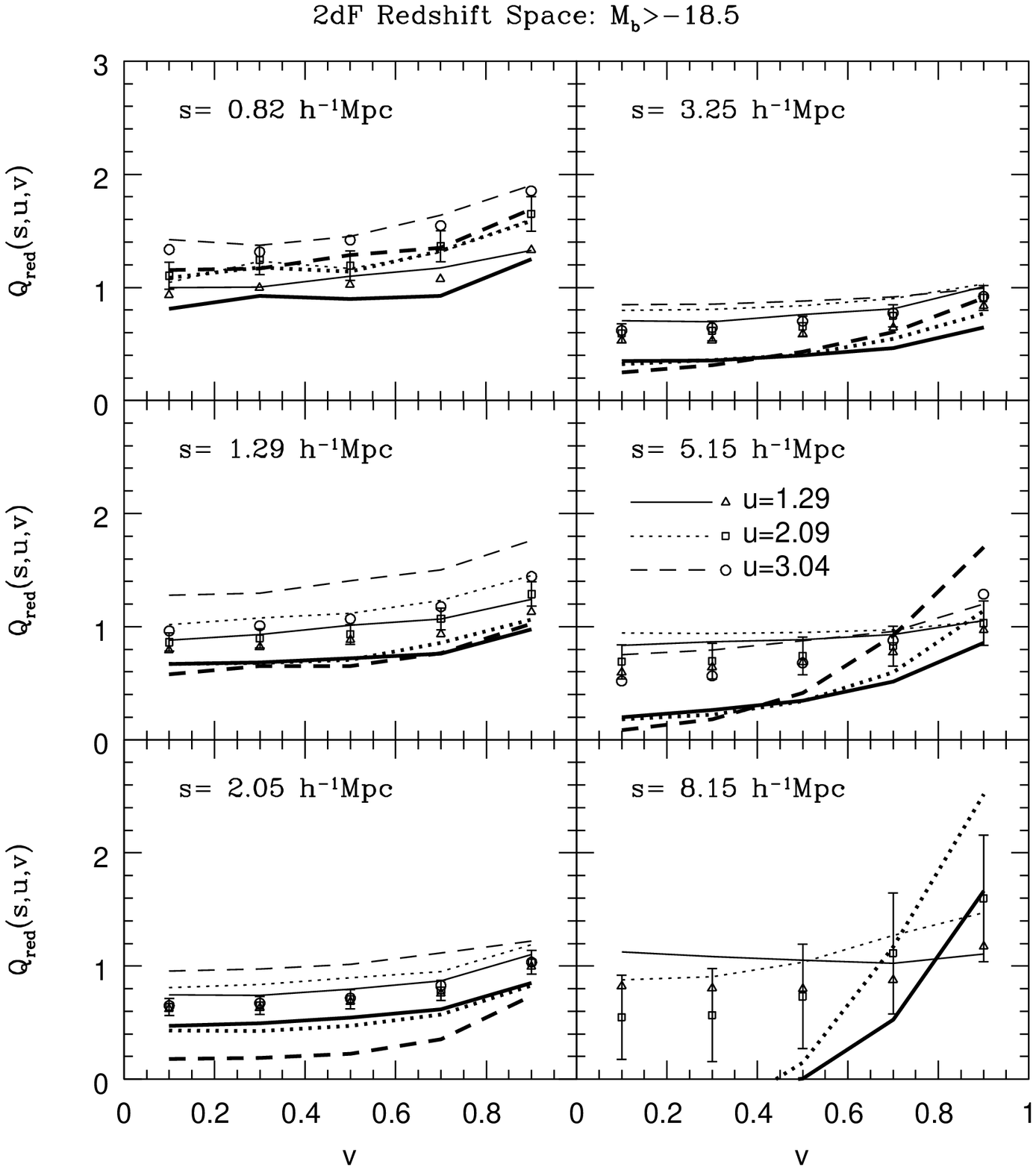}
\caption{The normalized 3PCF in redshift space $\Qsu$ of galaxies
with luminosity $M_b-5\log h> -18.5$ in the 2dFGRS survey.
The notations are the same as Fig.\ref{qred_1}.}
\label{qred_4}\end{figure}

\begin{figure}
\epsscale{1.0} \plotone{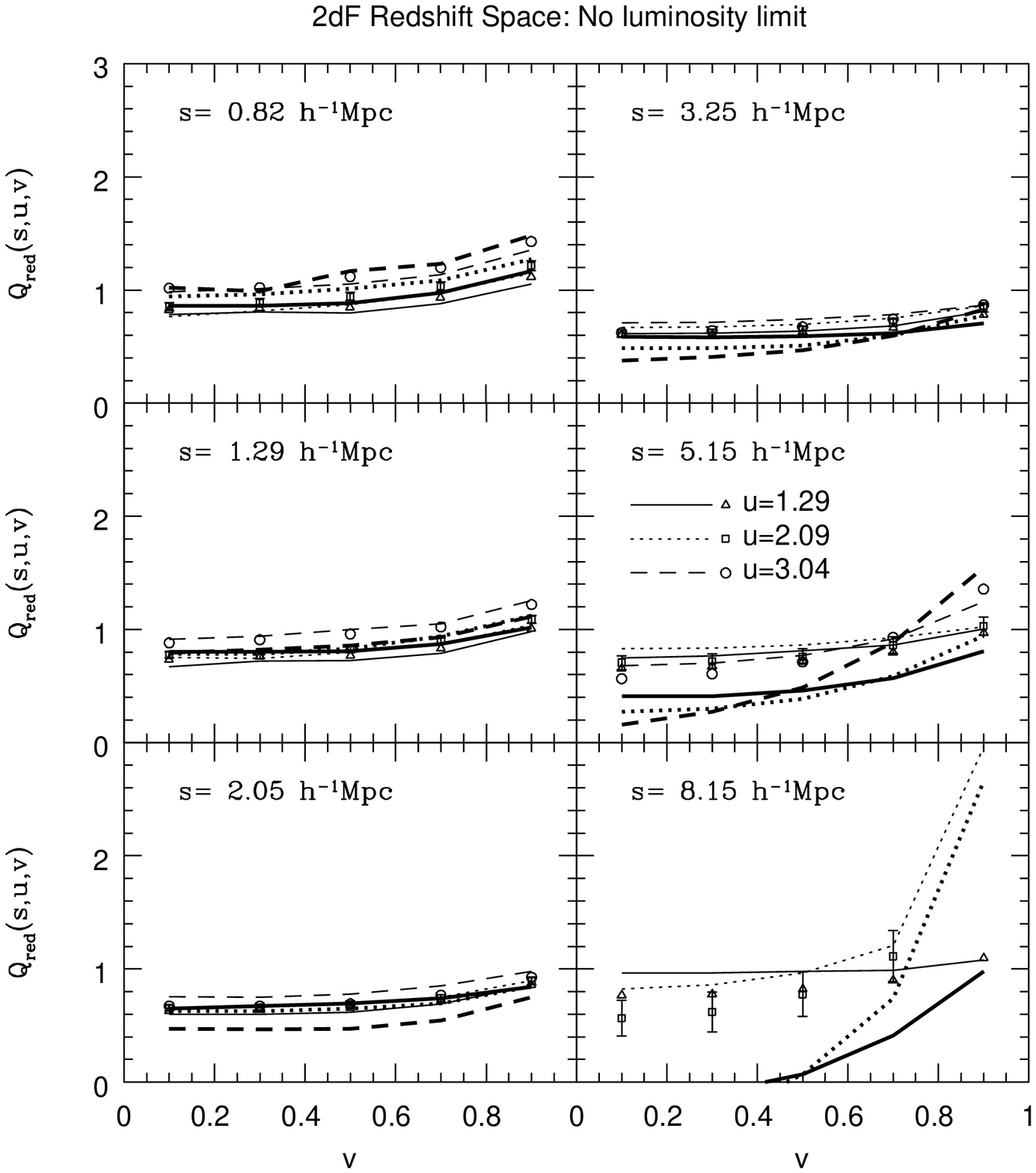}
\caption{The normalized 3PCF in redshift space $\Qsu$ of all galaxies
in the 2dFGRS survey (without luminosity selection).
The notations are the same as Fig.\ref{qred_1}.}
\label{qred_5}\end{figure}

\begin{figure}
\epsscale{1.0} \plotone{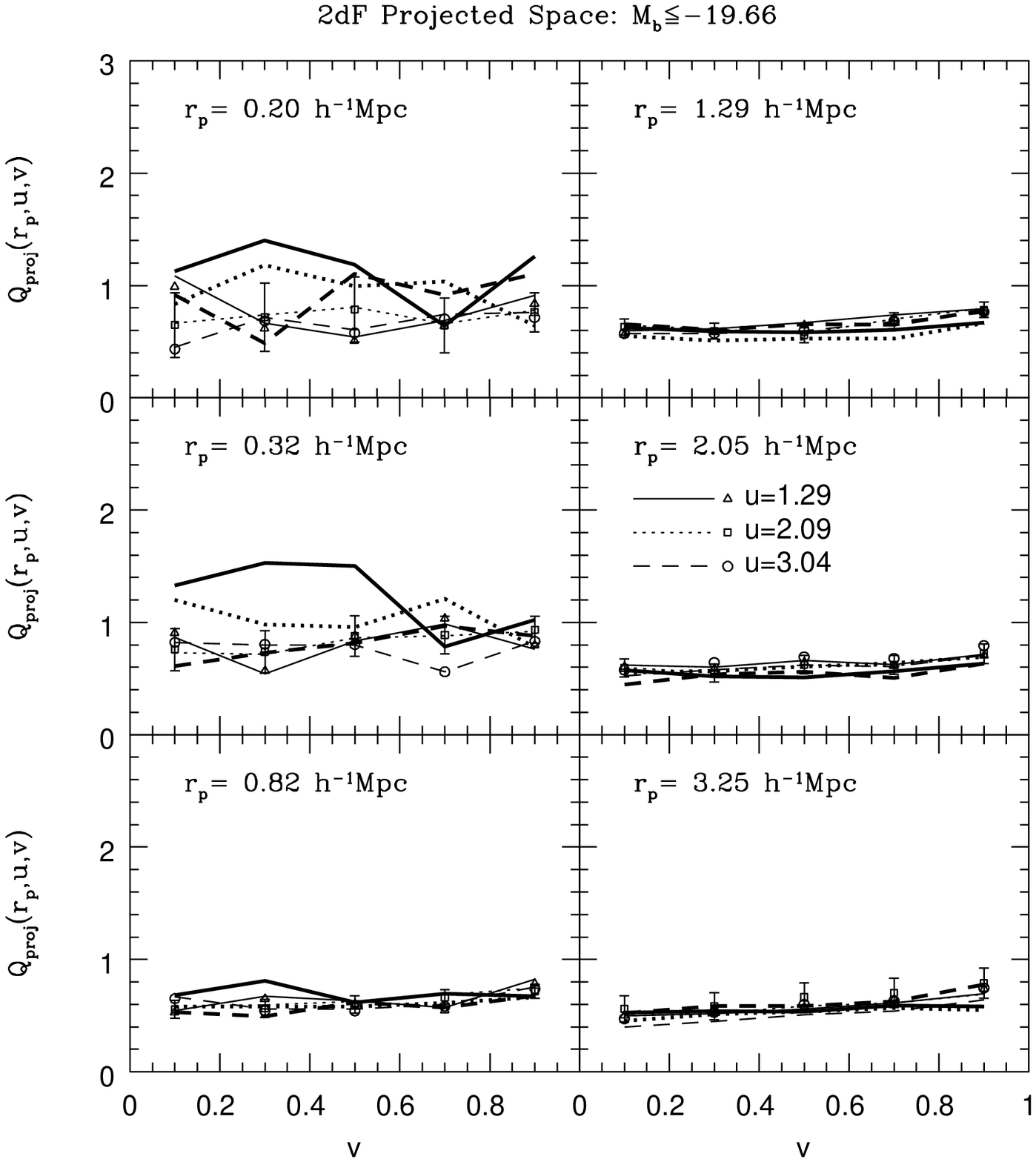}
\caption{ The normalized projected 3PCF $\Qrpu$ of galaxies with
luminosity $M_b-5\log h\le -19.66$ in the 2dFGRS survey. The results
for the south strip, the north strip, and the whole sample are plotted
with thick lines, thin lines, and symbols respectively. Different
lines and symbols are used for triangle configurations of different
$u$ as indicated on the figure. The errors are estimated by the
bootstrap resampling method. For clarity, the error bars are plotted
for the whole sample and $u=2$ only, but those for the other two
values of $u$ are very similar, and for the north or south strip are
about $1.4$ times larger.}
\label{qproj_1}\end{figure}

\begin{figure}
\epsscale{1.0} \plotone{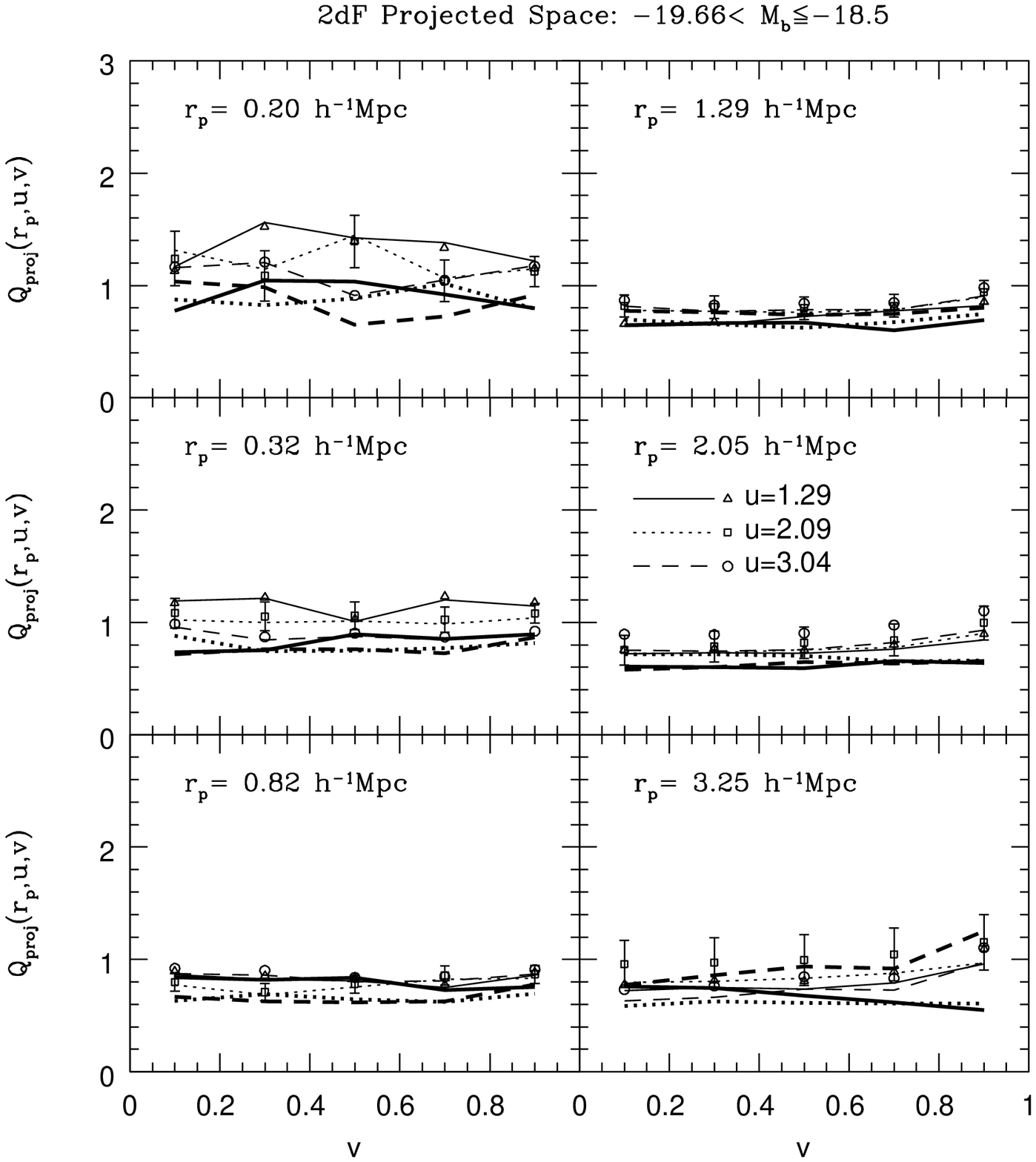}
\caption{The normalized projected 3PCF $\Qrpu$ of galaxies
with luminosity $-19.66<M_b-5\log h\le -18.5$ in the 2dFGRS survey.
The notations are the same as Fig.\ref{qproj_1}.}
\label{qproj_2}\end{figure}

\begin{figure}
\epsscale{1.0} \plotone{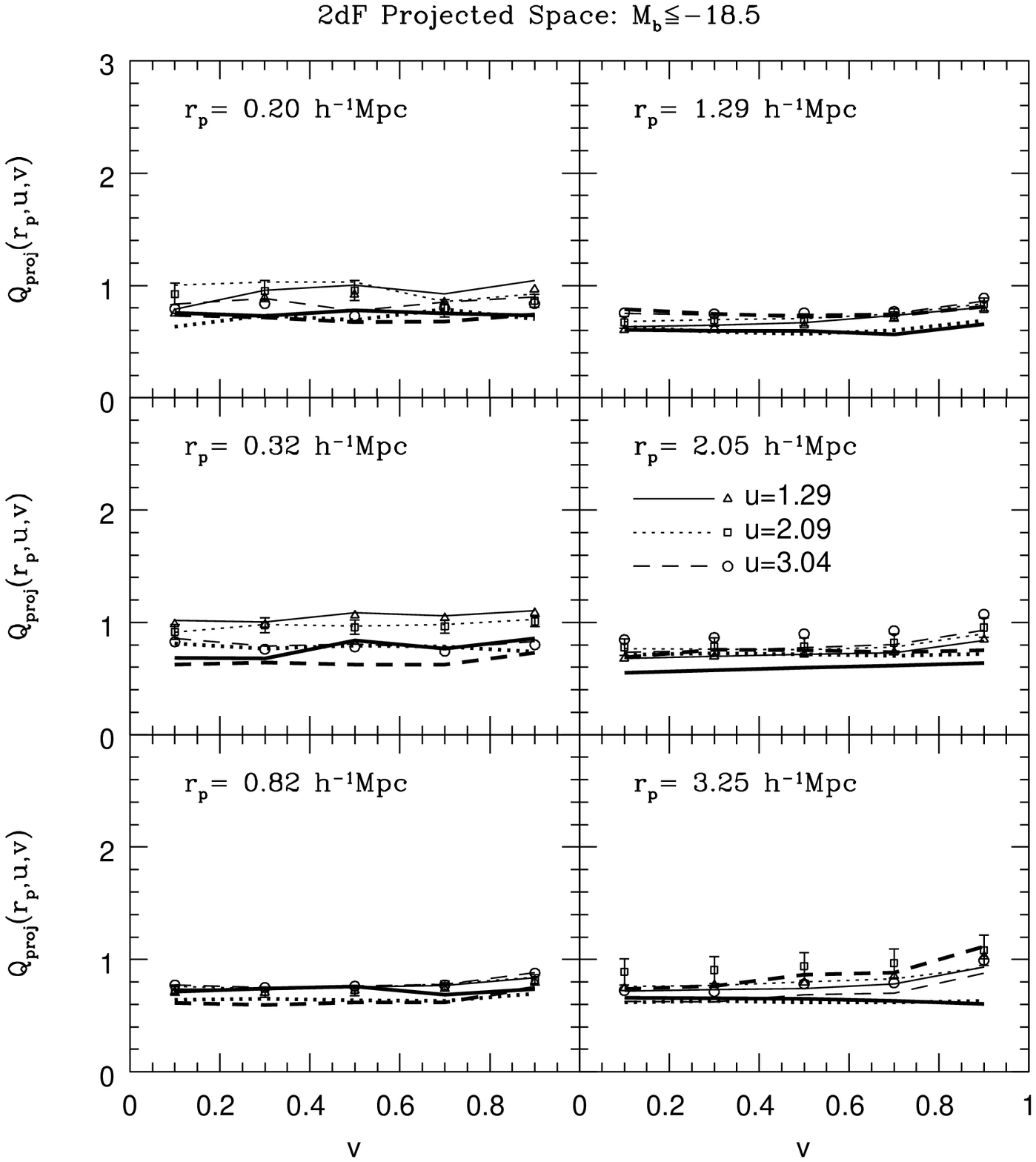}
\caption{The normalized projected 3PCF $\Qrpu$ of galaxies
with luminosity $M_b-5\log h\le -18.5$ in the 2dFGRS survey.
The notations are the same as Fig.\ref{qproj_1}.}
\label{qproj_3}\end{figure}

\begin{figure}
\epsscale{1.0} \plotone{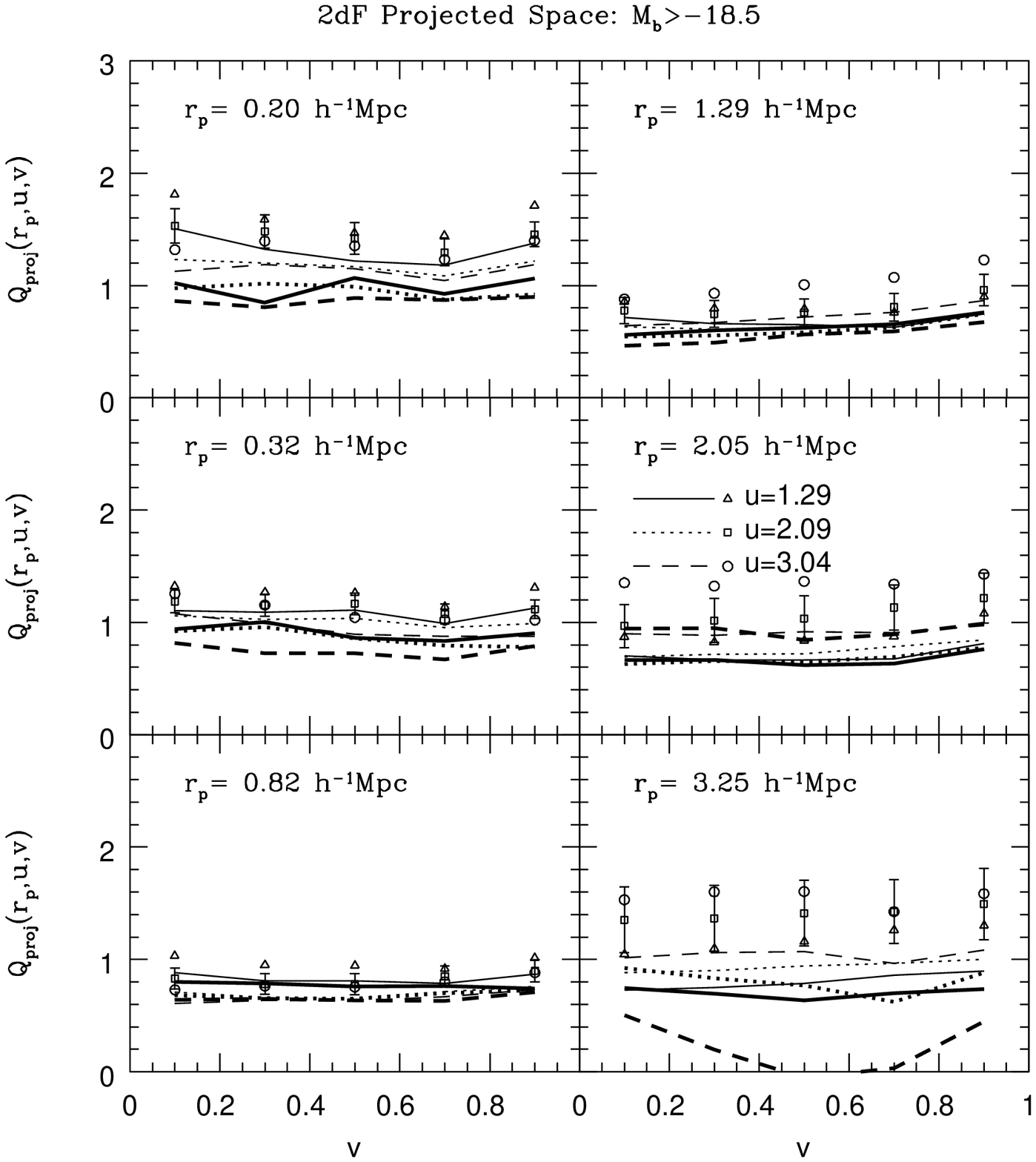}
\caption{The normalized projected 3PCF $\Qrpu$ of galaxies
with luminosity $M_b-5\log h> -18.5$ in the 2dFGRS survey.
The notations are the same as Fig.\ref{qproj_1}.}
\label{qproj_4}\end{figure}

\begin{figure}
\epsscale{1.0} \plotone{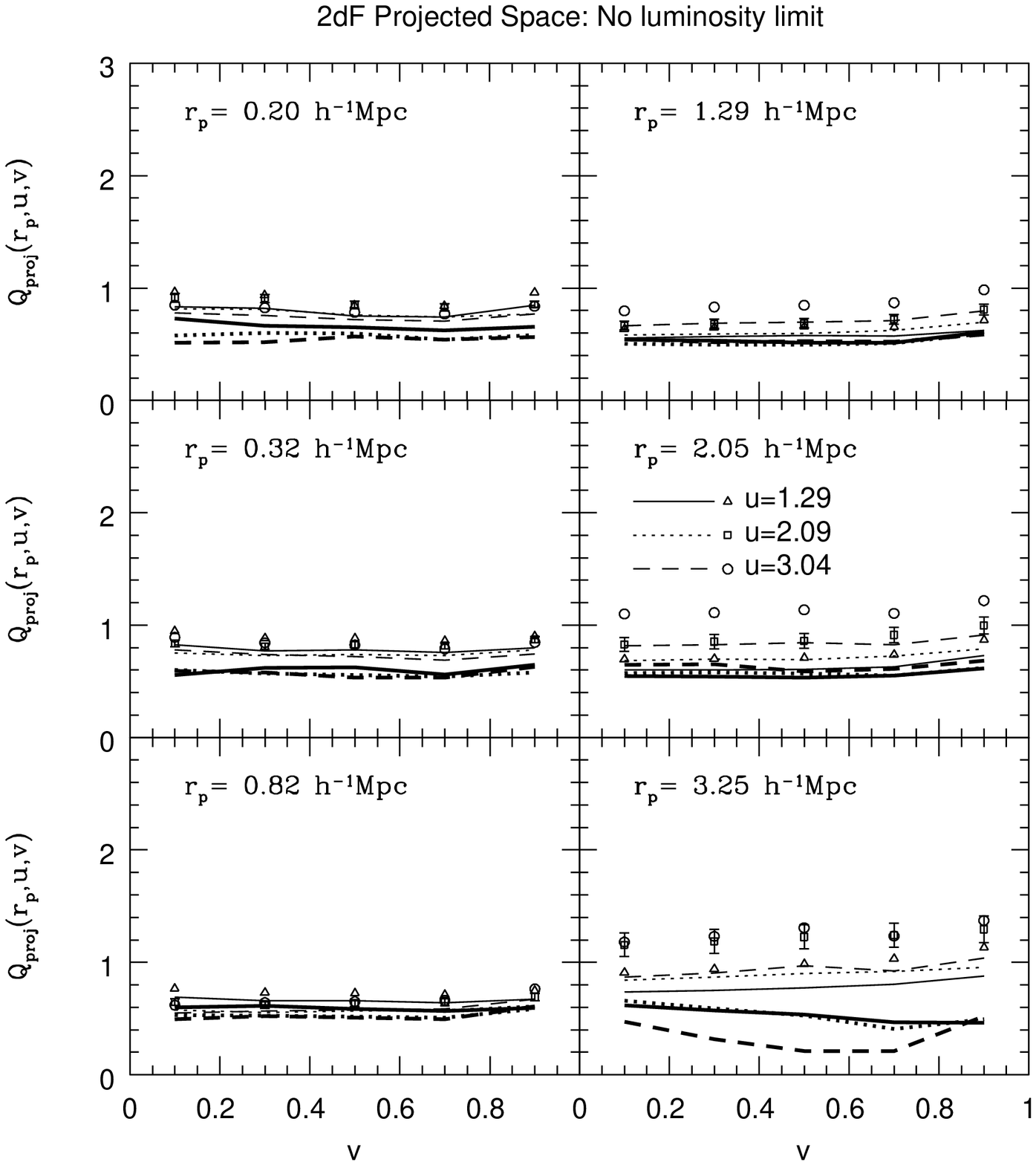}
\caption{The normalized projected 3PCF $\Qrpu$ of all galaxies
in the 2dFGRS survey (without luminosity selection).
The notations are the same as Fig.\ref{qproj_1}.}
\label{qproj_5}\end{figure}

We present the results of the 3PCF in redshift space $\Qsu$ and of the
projected 3PCF $\Qrpu$ in Figures \ref{qred_1} to \ref{qproj_5} for
the 2dFGRS survey. The errors of the Q-values are estimated by the
bootstrap resampling method. The large number of galaxies in the
2dFGRS survey allows us to look for a possible luminosity dependence
of $\Qsu$ and $\Qrpu$. We have selected five galaxy samples according
to luminosity, listed in Table 1. The samples are not completely
independent with significant overlaps between some of the
samples.

For $\Qsu$ the results are shown in Figures \ref{qred_1} to
\ref{qred_5}. As we can see , the 3PCF in redshift space is not
changing very much with $s$ or $u$, it increases somewhat with $ v$
for fixed $s$ and $u$.  For small $v$ $\Qsu$ is approximately constant
at a value of $\sim 0.6$, but it increases up to $\sim 1$, when $v
\sim 1$.

For the bright galaxies we find that $\Qsu$ decreases somewhat with
$s$, from $0.9$ at $s = 0.82 \mpc$ to $0.4$ at $s= 5.15 \mpc$. Changes
with $s$ are slightly reduced for the samples including fainter
galaxies.  For the faintest sample (IV), at small $s$ and $v$ $\Qsu$
is about $1.1$, and it decreases to $ \sim 0.7$ at $ s = 3.25
\mpc$. 

We find that $\Qsu$ is slightly larger for the fainter samples,
though the dependence on luminosity is rather weak. In fact, if the
errors are taken into account, this luminosity dependence is not
statistically significant. We also note that there is always some
difference between the north strip, the south strip, and the whole
sample, but generally within the $ 1\sigma$ error bars. This implies
that the bootstrap error used in this sample is a good indicator for
the error estimate. The results for the north and the south samples
are in good agreement for the galaxies brighter than $M_b-5\log h \leq
-18.5$. For the faint sample with $M_b-5\log h >
-18.5$, however, there is a significant difference between the north and
south subsamples. The main reason is that this sample 
covers only a small cosmic volume, so the
sample-to-sample difference (the cosmic variance) can be large. In
fact, even the 2PCFs of these subsamples are dramatically different
(see Figure \ref{xis}). Considering the fact that the bootstrap error
is not sufficient to fully account for the cosmic variance, one should
remain cautious about the result of the faintest sample
(IV). Nonetheless, from Figures \ref{qred_1} to \ref{qred_5} we
conclude that there is at best a slight dependence on luminosity 
in the sense that
the amplitude $\Qsu$ tends to be smaller for brighter galaxies.

The projected 3PCF in comparison shows a behavior which is somewhat
different. In the bright galaxy sample (Figure \ref{qproj_1}) $\Qrpu$
is about 0.7 at $r_p = 0.2 \mpc$, and it reaches down to $\Qrpu \simeq
0.5$ at $ r_p = 3.25 \mpc$ for small $v$, so the dependences on $r_p$
is quite mild. There is, however, a small but significant increase
with $v$.  Fainter galaxies show a similar weak dependence on $r_p$ and
$v$ (Figures \ref{qproj_2} and \ref{qproj_4}). But comparing different
samples, we find a trend that brighter galaxies have lower values of
$\Qrpu$. The $\Qrpu$ of the fainter samples (II and IV) is about 50\%
higher than that of the brightest sample of $M_b-5\log h < -19.6$. We
will discuss the implications for the bias parameters in \S 4.3.

The figures show that while the values of $\Qrpu$ 
are similar for the north and
south subsamples, the value for the total sample is larger than that
of either subsample. This looks a bit surprising at first glance. But
considering that the 2PCF of the north sample is almost 1.5 times
larger than that of the south sample on $\mpc$ scales, it is not
difficult to explain the behavior of $\Qrpu$ 
of the total sample and the two
subsamples. As an idealized example, we assume that the two subsamples
are well separated and have the same sample size, the same $\Qrpu$,
but the 2PCF of one sample is 1.5 times larger than that of the other. This
example is quite close to the real situation of the faintest
sample. It is not difficult to prove that the $\Qrpu$ of the total
sample is 1.4 times that of the subsamples. With this example, it
is easy to see that the amplitude $\Qrpu$ of the total sample is
larger than that of the subsamples for $r_p=1.29\sim 3.25 \mpc$ for
the faintest galaxies. This unusual behavior 
again can be attributed to
the fact that this sample surveys only a small volume of sky, so the
cosmic variance is large.

\subsection{Comparison with the results from the LCRS}

\begin{figure}
\epsscale{1.0} \plotone{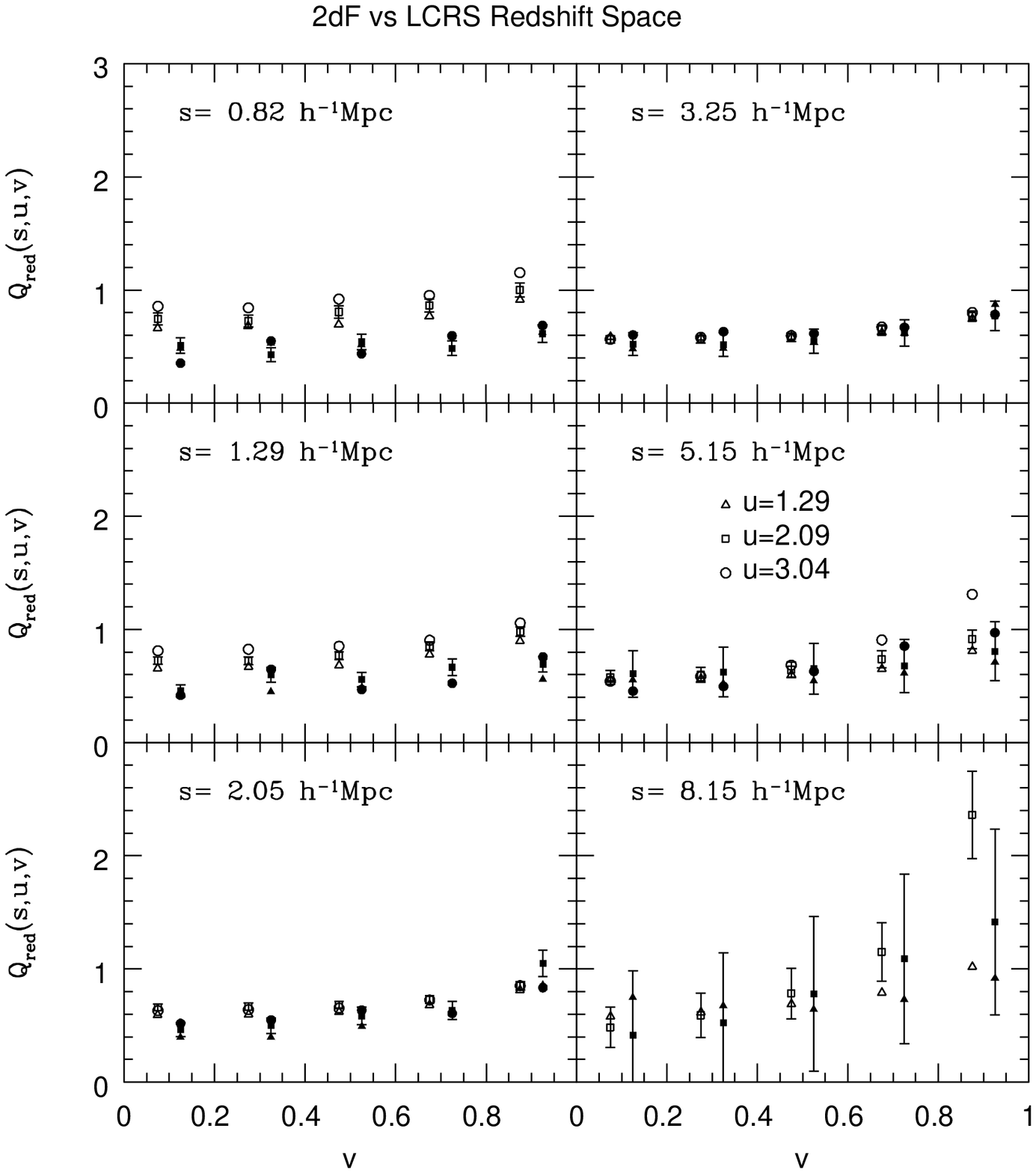}
\caption{Comparison of the normalized 3PCFs in redshift space $\Qsu$
measured from the 2dFGRS and from the Las Campanas Redshift Survey
(LCRS).  The data of the LCRS are taken from Jing \& B\"orner (1998) for
galaxies with luminosity in the R-band $M_R-5\log h\le -18.5$. For
comparison, we simply take our result in the 2dFGRS survey of galaxies with
$M_b-5\log h\le -18.5$, despite 
the fact that galaxies are selected in
different wavebands in the two surveys. The results of the LCRS are
plotted in solid symbols, and those of the 2dFGRS are in open symbols.
The errors are estimated by the bootstrap resampling method, and are
plotted for $u=2$ only. For clarity, the symbols are shifted by
$+0.05$ for the LCRS and by $-0.05$ for 2dFGRS along the horizontal
axis.}
\label{qred_2df_lcrs}\end{figure}

In Figure \ref{qred_2df_lcrs} we compare the normalized 3PCF in
redshift space $\Qsu$ of the 2dFGRS and Las Campanas surveys. The data
of the LCRS are taken from JB98 for galaxies with luminosities in the
R-band $M_R -5\log h \leq -18.5$. From the 2dFGRS we simply take our
result for the galaxies with $\mag \leq -18.5$, although we are aware
of the fact that the galaxies are selected in different wavebands in
the two surveys. There are subtle differences in the results which we
attribute to this choice of the observational bands, because $Q$
depends on luminosity weakly for $M> M^*$. For small values of $s\sim
1\mpc$, the 2dF catalog gives a slightly higher amplitude than the
LCRS galaxies. This could reflect the fact that the real space 2PCF of
the LCRS galaxies is higher than that of the APM galaxies on small
scales, as JMB98 pointed out.  Nevertheless, the $\Qsu$ values agree
very well between the two samples, especially on larger scales. The
2dF sample gives rise to a much smaller error, because of its large
sample size.

\begin{figure}
\epsscale{1.0} \plotone{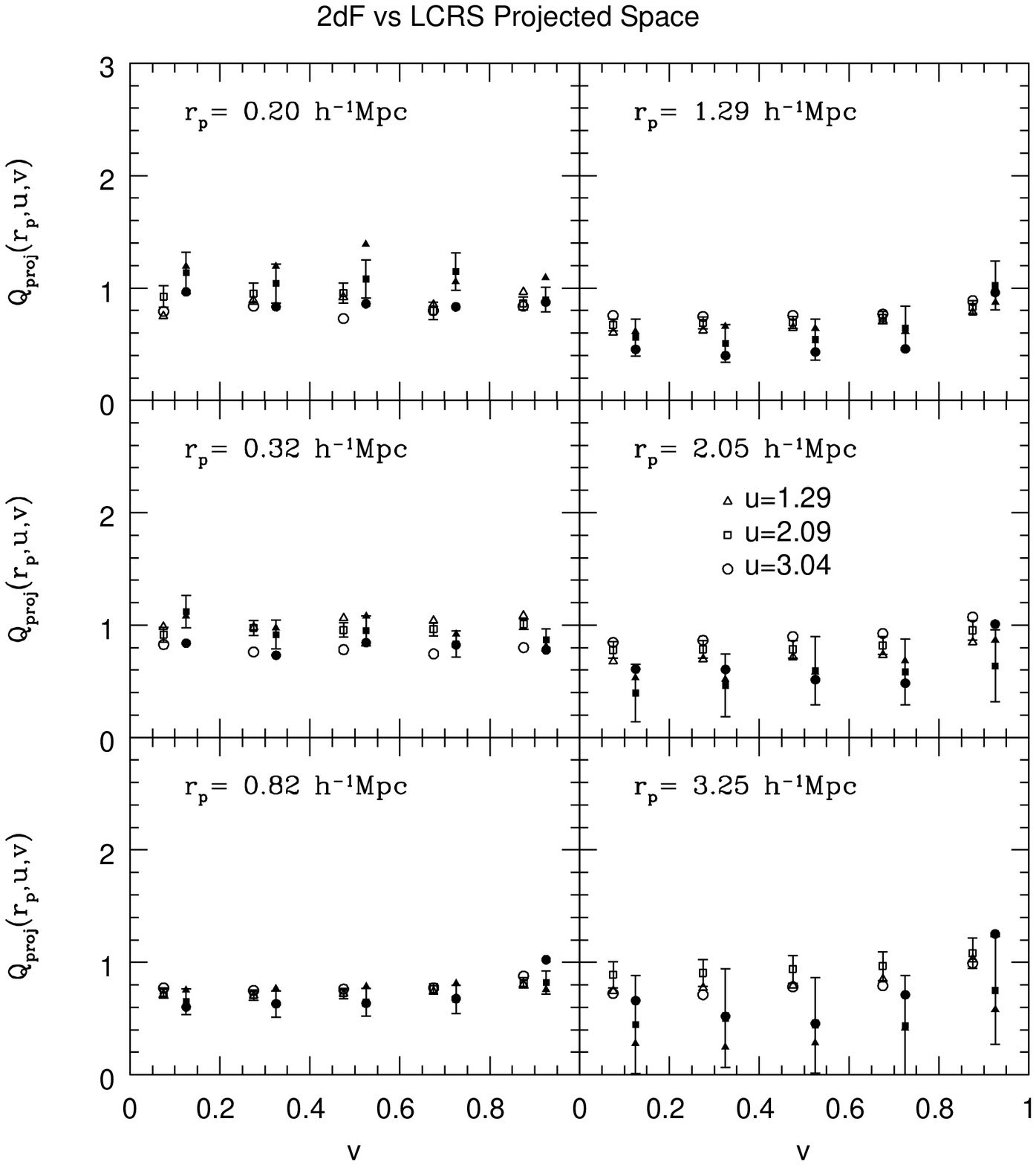}
\caption{Comparison of the normalized projected 3PCFs $\Qrpu$ measured
from the 2dFGRS and from the Las Campanas Redshift Survey (LCRS).  The
data of the LCRS are taken from Jing \& B\"orner (1998) for galaxies with
luminosity in the R-band $M_R-5\log h\le -18.5$. For comparison, we simply
take our result in the 2dFGRS survey of $M_b-5\log h\le -18.5$,
despite the fact that the galaxies are selected in 
different wavebands in the
two surveys. The results of the LCRS are plotted in solid symbols, and
those of the 2dFGRS are in open symbols.  The errors are estimated by
the bootstrap resampling method, and are plotted for $u=2$ only. For
clarity, the symbols are shifted by $+0.05$ for the LCRS and by
$-0.05$ for 2dFGRS along the horizontal axis.}
\label{qproj_2df_lcrs}\end{figure}

To compare the projected amplitudes $\Qrpu$, we display this quantity
for the two catalogs in Figure \ref{qproj_2df_lcrs}. Again the
agreement is quite satisfactory, especially when we take the larger
error bars for the LCRS result into consideration. However, the
systematic decrease with $r_p$ that can be read off for the mean
values of $\Qrpu$ for the LCRS data, is not present for the
2dFGRS. This is probably caused by the fact that the sky area of the
LCRS is much smaller than the 2dFGRS survey, so the mean value of the
LCRS is systematically underestimated. The 2dFGRS data also imply
that the real space 3PCF of galaxies on the small scales explored
here, does not deviate significantly from the hierarchical form
(P80), and that the fitting formula given in JB98 for the projected
$\Qrpu$ needs to be revised.  

In conclusion, our 2dFGRS results of $Q$, both in redshift space
and in projected space, are in good agreement with the 
results obtained by JB98 for the LCRS.

\subsection{Comparison with the dark matter distribution in 
the running power Cold Dark Matter model}

In this section, we compare the observational results with model
predictions.  Currently, the parameters of the Cold Dark Matter (CDM)
model have been determined pretty accurately by a combination of data
from WMAP, 2dFGRS, Lyman-$\alpha$ absorption systems, and
complementarily by many other observations \citep{spergel03}. We
choose the running power CDM model of Spergel et al. for comparison
with our statistical results, for this model matches most available
observations: The universe is flat with a density parameter
$\Omega_0=0.26$ and a cosmological constant $\lambda_0=0.74$. The
Hubble constant is $71 \kms {\rm Mpc}^{-1}$ and the baryonic density
parameter $\Omega_{0,b}=0.045$. The primordial density power spectrum
deviates slightly from the Zhe'dovich spectrum as $P(k)\propto (k {\rm
Mpc}/0.05)^{n(k)}$ with $n(k)=0.93-0.0165 \ln (k {\rm
Mpc}/0.05)$. Although there is no consensus about the necessity of
introducing the running power index $n(k)$ \citep[e.g. ][]
{seljak03,tegmark03}, we choose this model as a reasonable
approximation to the real situation.

Because the three-point correlation functions which we have measured,
are in the non-linear and quasilinear regimes, we use a N-body
simulation to make model predictions. The simulation has $512^3$
particles in a cubic box of $1024\mpc$, and is generated with our
$\pppm$ code \citep[see ][for the code]{js02}. To include the effect
of baryonic matter oscillations on large scale structures, the fitting
formula of \citet{eh99} for the transfer function is used to generate
the initial condition. Since the median redshift of the 2dFGRS is
$\sim 0.13$, we choose the simulation output at this redshift. We note
that the three-point correlation is quite sensitive to the presence of
very massive clusters, therefore a large simulation box like the one
used here is necessary. With a small box of $\ls 100 \mpc$ the
three-point correlation function may be underestimated severely.

Generally speaking, galaxies are biased tracers of the underlying
matter distribution in the Universe. A luminosity dependence of the
bias \citep{norberg02a} means that faint and bright galaxies trace the
matter distribution differently. It has become popular in recent years
to account for the bias of certain types of galaxies
phenomenologically with the so-called halo occupation model
\citep[e.g., ][ for an updated account of this model]{jmb98, seljak00,
ps00, sheth01, bw02, cs02, zehavi03, yang03a}. The three-point correlation
function of galaxies can also be modeled within this framework
\citep[][for a detailed account of this
modeling]{jb98,bw02,mf00,tj03}, though it seems difficult to account
for the two-point and three-point correlation functions in the LCRS
simultaneously with simple power-law occupation models
\citep{jb98}. Our accurate measurement of the 3PCF for the 2dFGRS and
its luminosity-dependence will certainly provide an even more
stringent constraint on the halo occupation models. It remains to be
seen, if the sophisticated model of Yang et al. (2003a,b) can explain
the results obtained in this paper. We want to investigate this issue
in a subsequent paper, and here we only compare with one model
prediction for the dark matter, in order to set a baseline quantifying
the difference in the normalized three point correlation function
between real galaxies and dark matter for the concordance CDM model.

\begin{figure}
\epsscale{1.0} \plotone{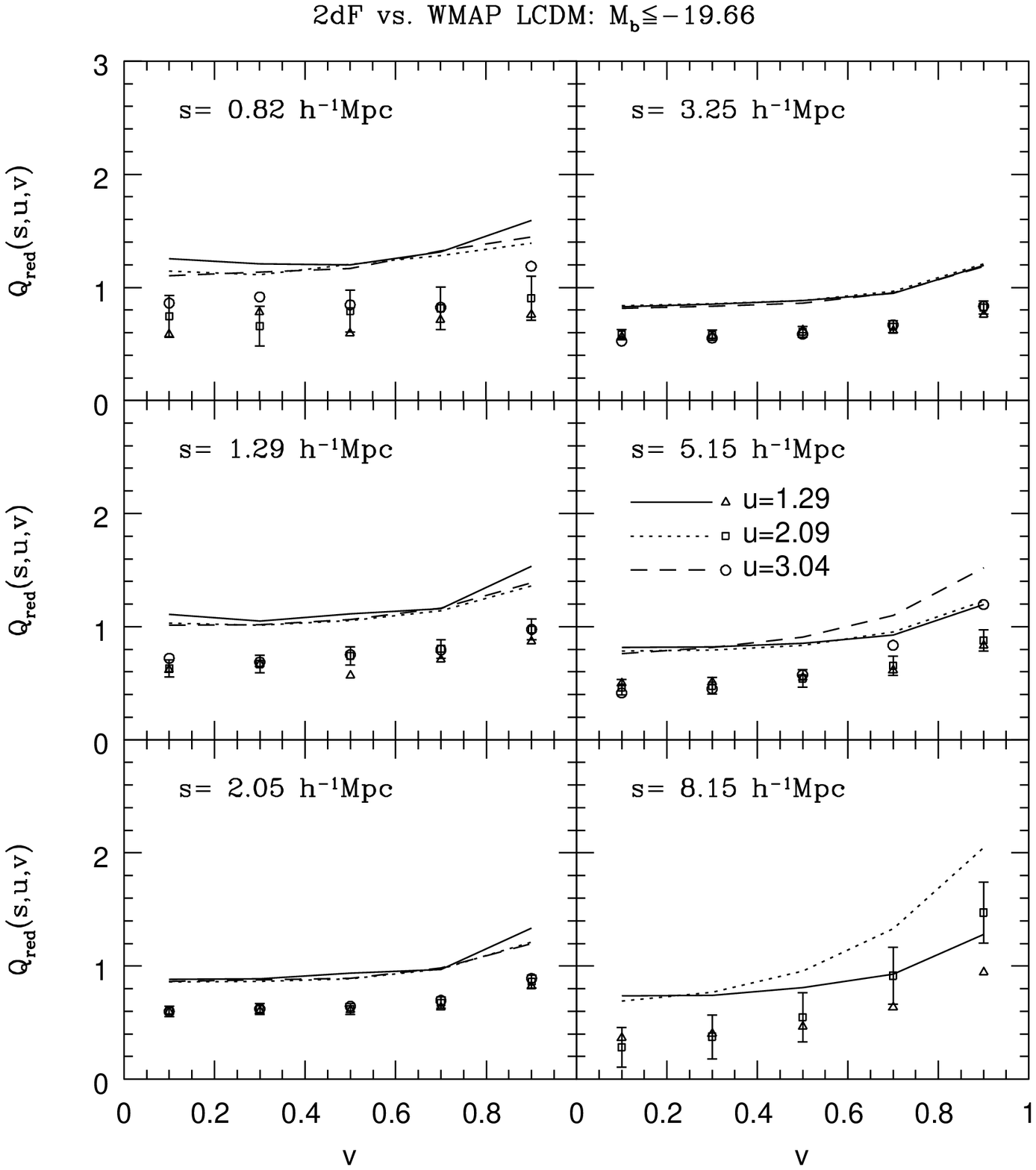}
\caption{Comparison of the normalized 3PCF of galaxies in redshift
space $\Qsu$ with the function predicted in the WMAP running power CDM
model. The observed data (sysmbols) are from the 2dFGRS survey for
$M_b-5\log h\le -19.66$, and the model prediction is for dark matter
(lines). The errors are estimated by the bootstrap resampling method,
and are plotted for $u=2$ only.  }
\label{model_qred_1}\end{figure}

\begin{figure}
\epsscale{1.0} \plotone{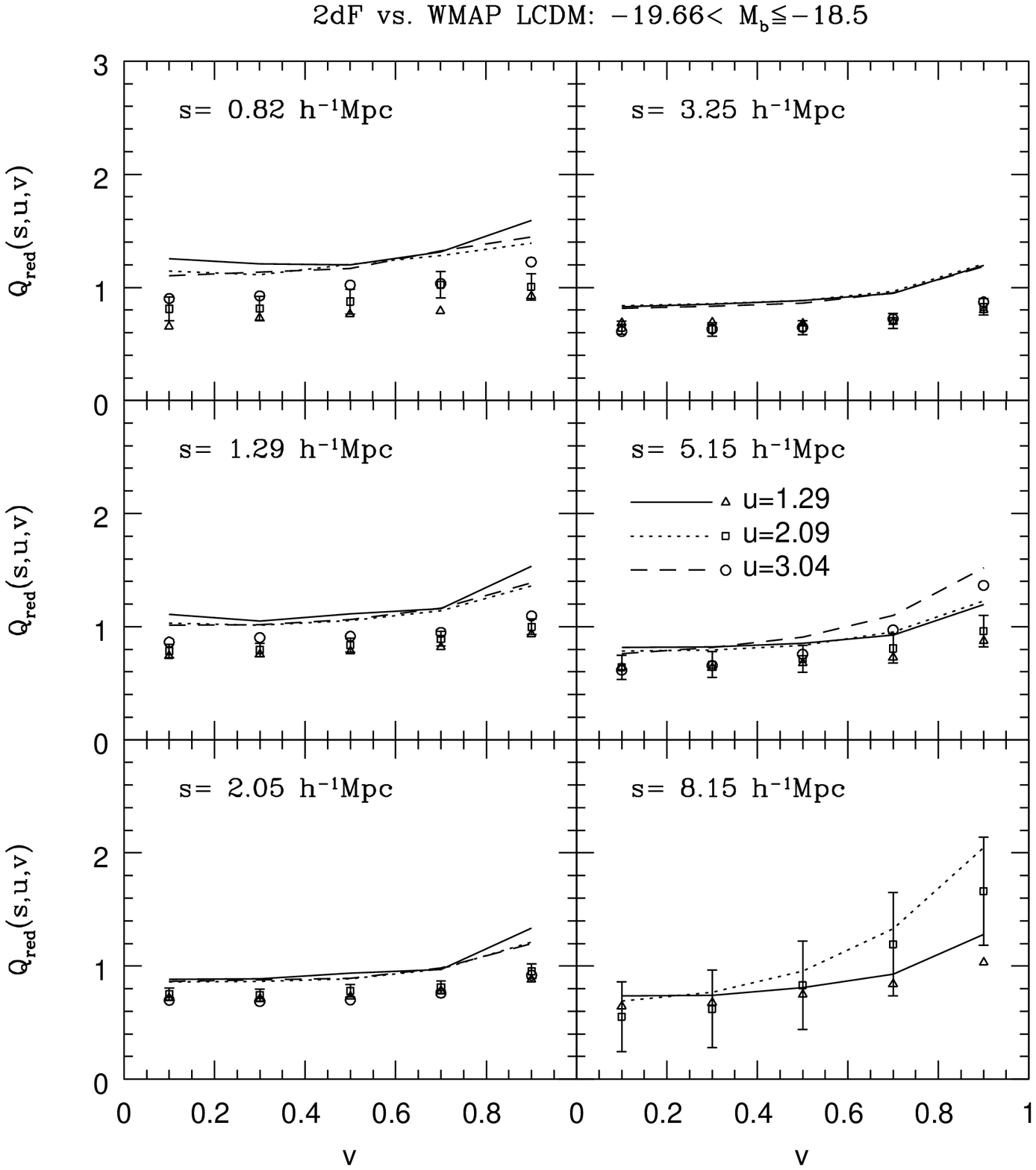}
\caption{Comparison of the normalized 3PCF of galaxies in redshift
space $\Qsu$ with the function predicted in the WMAP running power CDM
model. The observed data (sysmbols) are from the 2dFGRS survey for
$-19.66<M_b-5\log h\le -18.5$, and the model prediction is for dark
matter (lines). The errors are estimated by the bootstrap resampling
method, and are plotted for $u=2$ only.  }
\label{model_qred_2}\end{figure}

\begin{figure}
\epsscale{1.0} \plotone{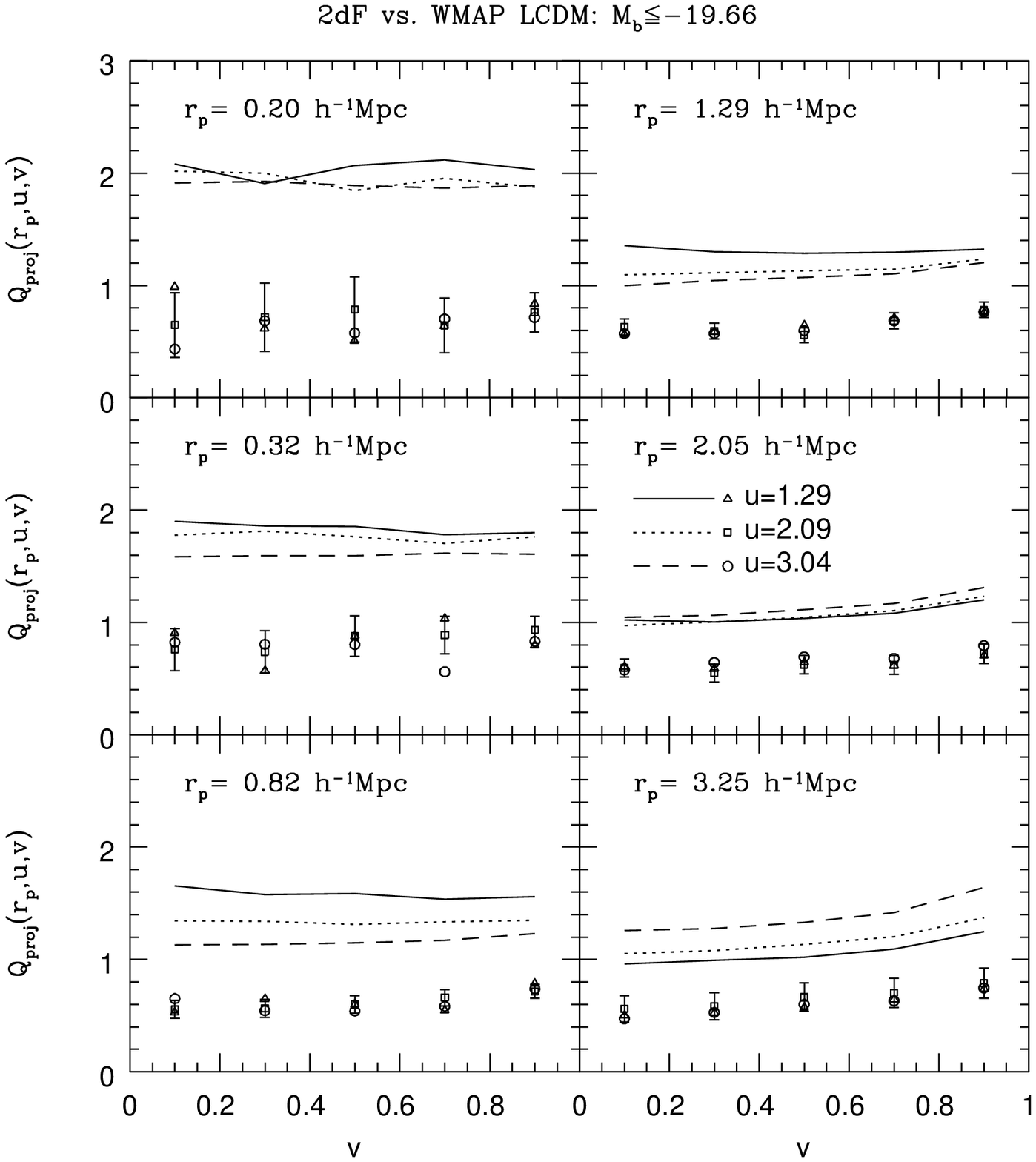}
\caption{Comparison of the normalized projected 3PCF of galaxies
$\Qrpu$ with the function predicted in the WMAP running power CDM model. The
observed data (sysmbols) are from the 2dFGRS survey for $M_b-5\log
h\le -19.66$, and the model prediction is for dark matter (lines). The
errors are estimated by the bootstrap resampling method, and are
plotted for $u=2$ only.  }
\label{model_qproj_1}\end{figure}

\begin{figure}
\epsscale{1.0} \plotone{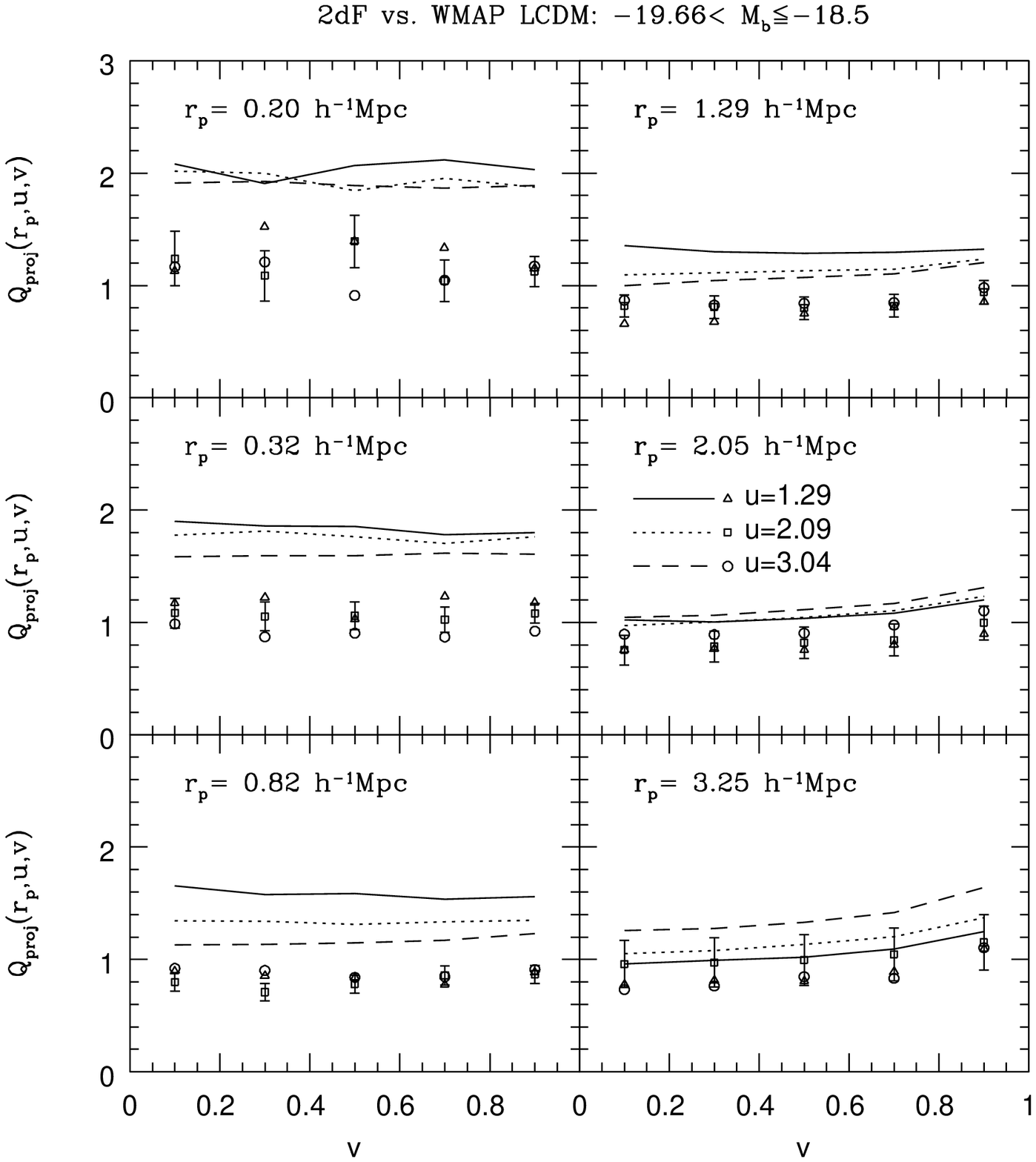}
\caption{Comparison of the normalized projected 3PCF of galaxies
$\Qrpu$ with the function predicted in the WMAP running power CDM model. The
observed data (sysmbols) are from the 2dFGRS survey for $-19.66<
M_b-5\log h\le -18.5$, and the model prediction is for dark matter
(lines). The errors are estimated by the bootstrap resampling method,
and are plotted for $u=2$ only.  }
\label{model_qproj_2}\end{figure}

The comparison between the 2dFGRS results and the model predictions is
displayed in Figures \ref{model_qred_1} to \ref{model_qproj_2}. Here
we consider only two luminosity subsamples. First, we find that the
qualitative features, such as the dependence on $v$ for fixed $s$ or
$r_p$, and $u$, the decrease of $Q$ with increasing values of $s$ or $
r_p$ are reproduced quite well by the DM simulations. For the luminous
sample (Sample I), the $Q$ values of the data set are generally lower
than the dark matter model predictions, up to a factor $1.5\sim
2$. For the less luminous sample (Sample II), the observed $Q$ values
also are smaller than those of the dark matter on small scales, but
the observed values and the model predictions agree at the values
$s\approx 8\mpc$ and $r_p=3.25\mpc$. Because the largest scales probed
here are expected to be linear or quasilinear scales, we expect the
linear bias model (eq.\ref{deltabias}) to hold on these scales. Our
result therefore tells us that on linear scales, the galaxies of
$-19.66<M_J-5\log h\le -18.5$ are approximately an unbiased tracer,
but the brightest galaxies of $M_J-5\log h\le -19.66$ have a bias
factor $\sim 1.5$.

Because the 2PCF of the galaxies of Sample II matches well the 2PCF of
the dark matter in the concordance WMAP model, and our 3PCF results
show that the galaxies of Sample II are unbiased on large, linear
scales, we find support for the density fluctuation normalization
$\sigma_8=0.84$ obtained by \citep{spergel03}. On the other hand, our
result shows that the three-point correlations $Q$ of galaxies are
lower on non-linear scales than the prediction of the WMAP concordance
model. Physical models, e.g. the halo occupation number model
\citep[e.g. ][]{yang03a} or the semi-analytical models of galaxy
formation \citep[e.g., ][]{kauffmann97} are needed to interpret the
observed small scale non-linear bias. We will pursue this in a future
paper.  The three-point correlation amplitudes of Sample III and
Sample V are very close to that of Sample II. The $Q$ of these samples
gradually conforms to the model prediction of the concordance model on
quasilinear scales $r\sim 5\mpc$. Our results are therefore consistent
with the analysis of Verde et al. who showed that the 2dFGRS galaxies
(without a luminosity classification) are an unbiased tracer of the
underlying matter on scales $5$ to $30\mpc$.

\section{Conclusion}

In a detailed analysis of the 3PCF for the 2dFGRS survey we have
accurately measured the 3PCF for galaxies of different luminosity. The
3PCF amplitudes ($\Qsu$ or $\Qrpu$) of galaxies generally decrease
with the increase of the triangle size and increase with the increase
of $v$, qualitatively in agreement with the predictions for the dark
matter clustering in popular hierarchical CDM models.  Some dependence
on luminosity is found, but not a strong effect, except for the
brightest galaxy sample which seems to have lower amplitudes of up to
$50 \%$.  Comparing with the previous study on the LCRS galaxies
(JB98), we find good agreement between the two studies, though the
results from the 2dFGRS are more accurate, since the 2dFGRS survey is
much larger than the LCRS survey.  The amplitudes in redshift space
$\Qsu$ are very similar, but the projected ones $\Qrpu$ show some
difference. It seems that the projected 3PCF from the LCRS is
systematically underestimated for $r_p$ in the range of a few $\mpc$,
because of the thin slice geometry of that survey. The dependence of
$\Qrpu$ on $r_p$ is much weaker in the 2dFGRS survey than in the LCRS
survey.

Comparing the measured 3PCF amplitudes ($\Qsu$ or $\Qrpu$) to the
prediction of a WMAP concordance model, we find that the measured
values are consistently lower than the predicted ones for dark
matter. As in the case of the LCRS about one-half of the predicted $Q$ value
provides a good description of $\Qrpu$ for the 2dFGRS data. As in JB98
for the LCRS data, we may state that the best fit DM model gives
higher values for the 3PCF than observed. This indicates that the
simple DM models must be refined, either by using more sophisticated
bias models, or a more sophisticated combination of model parameters.

The division of galaxies into luminosity classes reveals that the
brightest galaxies are biased even on large scales, while the galaxies
of sample II show a nonlinear bias on small scale, but appear unbiased
on linear scales.

\acknowledgments 

The work has made use of the data released by the 2dFGRS team, and the
software for generating mock samples provided by Peder Norberg and
Shaun Cole. We are grateful to Peder Norberg for his explanations on
how to use the masks and the softwares. We are grateful to Volker
M\"uller, Bob Nichol, Yasushi Suto for communicating their SDSS
results of the three-point correlations before publication. JYP would
like to thank the Max-Planck Institute f\"ur Astrophysik for its warm
hospitality during where the work was completed. The work is supported
in part by NKBRSF (G19990754), by NSFC (No.10125314), and by the
CAS-MPG exchange program.

\clearpage
 
\begin{table}
\caption{Samples selected according to luminosity}
\begin{center}
\begin{tabular}{cccrcccc}
\hline\hline Sample&$M_b-5\log
h$&South\tablenotemark{a}&North\tablenotemark{a}&Total\tablenotemark{a}\\
\hline I &$M_b-5\log h\le -19.66$&16702&11761&28463\\ II
&$-19.66<M_b-5\log h\le -18.5$&14247&11798&26045\\ III &$M_b-5\log
h\le -18.5$&30949&23559&54508\\ IV &$M_b-5\log h>
-18.5$&7930&6572&14502\\ V &no~limit&39208&30447&69655\\ \hline\hline
\tablenotetext{a}{Number of galaxies}
\end{tabular}
\end{center}
\end{table}

\end{document}